\documentclass{aafixed}
\usepackage{graphicx}
\usepackage{natbib}
\usepackage{rotating}
\usepackage[varg]{txfonts}
\usepackage{color}
\bibpunct{(}{)}{;}{a}{}{,}

\newcommand{\be}{\begin{equation}}
\newcommand{\ee}{\end{equation}}
\newcommand{\bea}{\begin{eqnarray}}
\newcommand{\eea}{\end{eqnarray}}

\newcommand{\bmp}{\begin{minipage}}
\newcommand{\emp}{\end{minipage}}

\newcommand{\bpic}{\begin{picture}}
\newcommand{\epic}{\end{picture}}

\newcommand{\mJy}{{\rm\thinspace mJy}}

\newcommand{\GHz}{{\rm\thinspace GHz}}

\newcommand{\km}{{\rm\thinspace km}}

\newcommand{\Mpc}{{\rm\thinspace Mpc}}

\newcommand{\s}{{\rm\thinspace s}}

\newcommand{\Hz}{{\rm\thinspace Hz}}

\newcommand{\W}{{\rm\thinspace W}}
\newcommand{\WHz}{\hbox{$\W \Hz^{-1}$}}

\newcommand{\kms}{\hbox{$\km\s^{-1}\,$}}

\newcommand{\kmpspMpc}{\hbox{$\km\s^{-1}\Mpc^{-1}\,$}}

\newcommand{\Msunyr}{{\rm\thinspace M_\odot \thinspace {\rm yr}^{-1}}}

\begin{document}
\title{Suppressed radio emission in supercluster galaxies: enhanced ram 
pressure in merging clusters?}
\titlerunning{Suppressed radio emission in supercluster galaxies}
\author{Jean-Christophe Mauduit\inst{1}, Gary A. Mamon\inst{1,2}}

\authorrunning{Mauduit and Mamon}

\offprints{Gary Mamon, email: gam AT iap.fr}

\institute{
Institut d'Astrophysique de Paris (UMR 7095: CNRS and Univ. Pierre \& Marie Curie), 
98 bis Bd. Arago, F--75014 Paris, France \and 
GEPI (UMR 8111: CNRS and Univ. Denis Diderot), Observatoire de Paris, 
F--92195 Meudon, C\'edex, France}

\date{Received 25 April 2007 / Accepted 31 August 2007}

\abstract
{Continuum radio emission of galaxies is related to AGN activity and to
  starbursts, both of which require a supply of gas, respectively  
to the central black hole
  and to molecular clouds.}
{The environmental influence on the 21cm (1.4 GHz) 
continuum radio emission of galaxies
is analyzed in a 600 deg$^2$ region of the local Universe containing the
Shapley Supercluster (SSC), whose core is thought to be the site of
cluster-cluster merging.}
{Galaxies in the FLASH and 6dFGS optical/NIR redshift
surveys are cross-identified with NVSS radio sources, selected in a subsample
doubly complete in volume
and luminosity, and classified as
starbursts or AGN according to their radio luminosity.
We study radio luminosities as well as 
radio loudness (luminosities normalized by stellar mass) ${\cal R_K}$.
Environmental effects are studied through a smoothed density field
(normalized to that obtained from random catalogs 
with the same survey edges and redshift selection function) and through the
projected distance to the nearest cluster (in units of its virial radius,
whose relation to the aperture velocity dispersion is quantified).
}
{The fraction of high ${\cal R}_K$ galaxies in the dense 10 Mpc
  Abell~3558  
cluster complex at
 the core of the SSC (SSC-CR) is half as large than elsewhere.
Moreover, radio loudness in the SSC-CR 
is anti-correlated with the density of the
large-scale environment and correlated with clustercentric radius:
central brightest cluster galaxies (BCGs) in the SSC-CR 
are an order of
magnitude 
less radio-loud than BCGs elsewhere, with signs of suppressed radio loudness
also present beyond the BCGs, out to at least $0.3\,r_{200}$.
The gradual suppression of radio loudness from inner cluster
regions to the cluster centers highlights a significant correlation of radio
loudness with clustercentric radius, not seen outside the SSC-CR.
This correlation is nearly as strong as the tight correlation of $K$-band
luminosity, $L_K$, with clustercentric radius ($K$-luminosity segregation), 
inside the SSC-CR, with a mild $K$-luminosity segregation outside the SSC-CR.
}
{
The suppression of radio loudness in SSC-CR BCGs 
can be attributed to
cluster-cluster mergers that destroy the cool core and thus the supply
  of gas to 
  the central AGN.
We analytically demonstrate that 
the low radio loudness of non-BCG galaxies within SSC-CR clusters
cannot be explained by
direct major galaxy mergers or rapid galaxy  flyby collisions,
but by the loss of gas supply through
the enhanced ram pressure felt when these galaxies 
cross the shock front between the two merging clusters and
are afterwards subjected to the stronger wind from the second
cluster.}


\maketitle

\section{Introduction}
\label{sec:intro}

Galaxies are not isolated objects
and evolve in tight relation to the structure they belong to. These
environmental effects are expected to hold even more strongly for galaxies with important
emission in the radio continuum, i.e. radio galaxies. Such galaxies range from the less radio 
luminous ones, star-forming galaxies (hereafter SFGs), to the highest radio-luminous ones, 
galaxies with an active galactic nucleus (AGN). 
However, the exact role of the environment on galaxy activity -- starburst or 
nuclear -- is still not clear. 

The first environmental studies of radio galaxies were based upon small
samples of galaxies distributed over the entire sky to derive their
clustering properties. 
In a pioneering study, \cite{Dressler85} showed that low luminosity radio
galaxies (i.e. SFGs) avoid dense environments. 
\cite{Peacock91} found that radio galaxies are more strongly clustered than
individual galaxies but less clustered than clusters of galaxies. 
Galaxies of intermediate
radio luminosity tend to lie in poor groups \citep{Bahcall92}. The 
most powerful radio sources appear to favor galaxy groups and poor 
clusters \citep{Prestage88,Hill91}. 

In recent years, the advent of large radio and optical catalogs of galaxies has
permitted a better statistical characterization of these trends.
Large-scale studies have matched the angular positions of radio
galaxies in the NRAO VLA Sky Survey (NVSS, \citealt{Condon98}) and Faint
Images of the Radio Sky at Twenty Centimeters (FIRST, \citealt{Becker95})
with the redshift-space positions of optical galaxies in the 2dF Galaxy
Redshift Survey (2dFGRS, \citealt{Colless01}) or the Sloan Digitized Sky
Survey (SDSS, \citealt{Strauss02}). 
Many of the previous trends noted above have been confirmed and developed.
\cite{Magliocchetti+04} showed that 
AGN appear to be strongly clustered and find no significant differences in the 
clustering of faint and luminous radio galaxies. 
\cite{Best04} 
found that the fraction of SFGs 
(defined as galaxies of low radio luminosity)
decreases with the density of the environment and that AGN are preferentially located 
in groups and poor to moderate richness clusters and avoid regions of very low density. 
In compact groups of galaxies, the SFG fraction increases drastically from the core to 
the halo while AGN seem to always lie in the cores \citep{Coziol98}. 
In clusters too, SFGs are broadly distributed, whereas AGN are centrally 
concentrated \citep{Miller02}. 
However, AGN that are brightest cluster galaxies (BCGs) have the same
  distribution of radio luminosities as non-BCGs \citep{Best+07}.
All these studies seem to indicate that 
radio galaxies, in general, prefer environments of intermediate density, with different 
behaviors at both ends of the radio luminosity distribution, 
the lowest radio emission avoiding dense environments and the most powerful 
ones avoiding very low density regions. 
%

Since the fraction of radio galaxies (both SFGs and AGN) seem to depend more
on the large-scale environment than on the small-scale one \citep{Best04}, it
is worth investigating the properties of radio galaxies on the scales of
superclusters. Indeed, these superstructures 
contain a wide variety of environments, 
from small galaxy groups to big clusters.
 It is be therefore interesting to mix the two 
approaches (the large-scale and local environment) over an area containing 
diverse environments, from empty regions (voids) to the densest ones 
(a supercluster).

In the present work, we study the effect of the supercluster environment on
the radio luminosity and $K$-band normalized radio loudness of galaxies.
For this, we
analyze the 
deepest, homogeneous redshift sample of radio-emitting
galaxies covering the Shapley
  Supercluster and its surroundings in a region
$70^\circ \times 10^\circ$.

The Shapley Supercluster (SSC) is the densest region in the local universe 
\citep{Ray91,Fabian91}. The SSC and some of its clusters have been studied in 
detail over the years, mainly by \cite{Quin95,Quin00}, \cite{Bardelli00,Bardelli01} 
and \cite{Drink04} and its radio properties were first investigated by 
\cite{Venturi97}. Moreover, this region is interesting because it hosts
a variety of environments. In particular, its central, 10 Mpc
  radius region,
  known as Shapley~8 or the \object{Abell~3558} (A3558)
cluster complex (hereafter SSC-CR
  following  
the nomenclature of \citealt{Quin00}),
is thought to be globally undergoing gravitational collapse
  \citep{Reisen00}, with its central clusters, Abell~3556, 3558 and 3562
  possibly merging today with one another and with smaller galaxy 
groups \citep{Bardelli00,Venturi00}.

Several authors recently pointed out that cluster-cluster merging can 
have dramatic consequences on radio galaxies and might be the key to the general 
trends observed. But these consequences remain uncertain. 
For example, \cite{Owen99} 
and \cite{Miller03} found an increase of star-formation and AGN 
activity that might be enhanced in several Abell clusters
by an ongoing merger with
another cluster.
The dense regions of superclusters are expected to be an ideal laboratory for 
studying colliding clusters.
In the SSC-CR, \cite{Venturi00} found that the A3558 
cluster presents a deficit 
of radio galaxies. On the other hand, another cluster complex outside the SSC-CR,
centered around \object{Abell~3528} (A3528), appears to have different
physical 
characteristics, as 
\cite{Venturi01} found the A3528 complex to be active at radio wavelengths, although its 
radio luminosity function turns up being similar with the generic one derived 
by \citet{Ledlow96}.
\cite{Bardelli01} argued that this complex was possibly in a
pre-merger state.

Cross-identification with NVSS radio sources gives us access, not only to the 
radio luminosity but also to the radio-loudness (here defined as power normalized 
to stellar mass -- from $K$-band luminosity), a variable that has been 
generally overlooked in previous studies. 

The merging of the optical, near-infrared (NIR) and radio catalogs is described in
Sect.~\ref{sec:catalogs}. 
In Sect.~\ref{sec:radiofracs}, we compare the distributions of radio
  luminosity and of radio loudness between the A3558 complex, the A3528
  complex and the remaining FLASH region.
We study
how the radio power and loudness are modulated by the density of the
large-scale environment in Sect.~\ref{sec:envmt} and with the
  clustercentric 
radius in Sect.~\ref{sec:cluster}.
We compare our results with other studies in Sect.~\ref{sec:compare}, 
and discuss them in Sect.~\ref{sec:process},
making
qualitative and semi-quantitative predictions of the effects of galaxy mergers, 
rapid flyby collisions and altered ram pressure stripping on galaxies within colliding 
and merging clusters.
In two appendices, we compute the relation between aperture velocity
  dispersion and
  virial radius for a \citeauthor*{NFW96} (\citeyear{NFW96}, hereafter NFW) model
 and the rate of direct galaxy mergers in overlapping
  colliding clusters of equal mass.
A summary is provided in Sect.~\ref{sec:sum}.
Throughout this paper the values adopted for the cosmological parameters are 
$\Omega_m=0.3, \Omega_{\Lambda}=0.7$ and a Hubble constant $H_0=70$ \kmpspMpc.

\section{Catalogs}
\label{sec:catalogs}

\subsection{Redshift catalogs}
\label{subsec:optcats}

We cross-correlate radio-detected galaxies from the NVSS radio survey with galaxies from 
two different optical/NIR surveys: the FLAir Shapley-Hydra survey (FLASH) and
the Six degree Field Galaxy Survey (6dFGS).

The FLASH survey\footnote{The FLASH survey catalog was obtained via the VizieR 
On-line Data Catalog: J/MNRAS/339/652.} \citep{Kaldare03}, is a redshift survey 
containing 3141 galaxies with measured redshifts in a $70^{\circ} \times 10^{\circ}$ 
strip, aligned in galactic coordinates ($260^\circ < \ell < 330^\circ$ and 
$25^\circ<b<35^\circ$), extending from the Shapley Supercluster (SSC) to the Hydra 
cluster, covering a solid angle of $605\,\deg^2 = 0.184\,\rm sr$. Its input catalog 
is the Hydra-Centaurus Catalogue of \cite{Raychaudhury90}, a photometric catalog down 
to $b_J = 17$, compiled by scanning UKST Southern Sky Survey plates with the Automated 
Photographic Measuring (APM) facility in Cambridge. Absolute positions are accurate 
to $\sim 1''$. The FLASH survey magnitude limit, after correction for Galactic extinction, 
is $b_J = 16.7$. Radial velocities were obtained with the FLAIR-II spectrograph mounted 
at the UKST. Additional data from the literature (NED, ZCAT) was incorporated into the 
original FLAIR redshift sample.
The rms error in the redshifts is $95 \, \rm km \, s^{-1}$. 
\cite{Kaldare03} assert 
that the only redshift bias in their sample is due to magnitude-dependent incompleteness, 
which is $50\%$ at the $b_J = 16.7$ survey limit. The overall spectroscopic completeness 
of the catalog is of 68\% with redshifts for 3141 of the 4613 galaxies in the catalog. 
The catalog median depth is $\sim 10\,000 \kms$.

We combine this survey with the 6dFGS survey \citep{Jones+04} Data Release~2.
\footnote{The 6dFGS Second Data Release is publicly available from the 6dFGS website at
http://www.mso.anu.edu.au/6dFGS/.}
As its principal input catalog, 6dFGS uses the $K$-band galaxies from the Two Micron 
All Sky Survey (2MASS, \citealt{Jarrett00}) and includes all galaxies brighter than
$K_{\rm tot} = 12.75$ (corresponding to roughly the same $b_J$ limit as that of the 
FLASH survey). 
Contrary to optically selected galaxies, $K$-band selected galaxy samples 
are not biased towards recent star formation activity, and the low extinction in the 
$K$-band ensures a deeper view through the interstellar dust of the observed galaxies 
(as well as through our Milky Way). 6dFGS velocities were obtained with the robotized 
6dF spectrograph mounted at the UKST (in replacement of FLAIR II). We retrieve the 
6dFGS galaxies with a redshift quality factor of 4
(reliable redshift) 
in the FLASH area. 
The median depth of the catalog is $\langle cz \rangle = 16\,008 \kms$ 
and the error on $Q=4$ velocities is $46 \, \rm km \, s^{-1}$.
In this catalog, a galaxy may appear several times due to multiple observations. 
In order to have a unique sample of 6dFGS galaxies, we replace
 multiple redshift measurements by their mean.

We merge these two redshift catalogs 
by cross-identifying them
with a search radius of $6''$, 
keeping the (more precise) 6dFGS velocity for matching pairs and 
adding non-matched FLASH galaxies to our catalog.
As this work reached completion, 
a new set of velocities has been published by \cite{Proust+06}, which goes to
fainter magnitudes in some regions. However, we chose not to use 
these velocities because of the inhomogeneous selection and smaller solid
angle of this data set. 
In the end, our ``6dFGS+FLASH catalog'' consists of 5132
galaxies with redshifts in the FLASH region.

\subsection{Radio catalog}
\label{subsec:radiocats}

The NVSS is a radio continuum survey at frequency $1.4\,\rm GHz$ (21cm) covering the sky 
north of $-40^\circ$, thus fully overlapping the FLASH area.\footnote{A text version 
of the NVSS catalog, NVSSCatalog.text, is available by ftp at nvss.cv.nrao.edu. 
It was generated with NVSSlist version 2.17 (August 2001 B. Cotton) on the entire FITS 
database.} 
The source catalog \citep{Condon98} contains $\sim~1.8 \times 10^6$ entries.
It has an 
angular resolution of $45''$ (FWHM).
\citeauthor{Condon98} have estimated the differential completeness to be 99\%
at flux $S_{1.4\GHz} = 3.5 \mJy$, and from their analysis we infer a
differential completeness of 75\% at 2.8 mJy.
Selecting the sources in the FLASH area, 
we retrieve $\sim 68\,000$ radio sources. Although the FIRST survey is a similar 
survey with a better angular resolution of $5''$ and a completeness of $95\%$ 
at $S_{1.4\GHz} > 2$ mJy, it does not cover further South than $-10^\circ$ in declination, 
which is not enough to reach the SSC region.

\subsection{Merged radio-NIR galaxy sample and radio galaxy classification}
\label{subsec:xid}

\begin{figure}
\hspace{5cm}
\centering
\includegraphics[width=8.5cm]{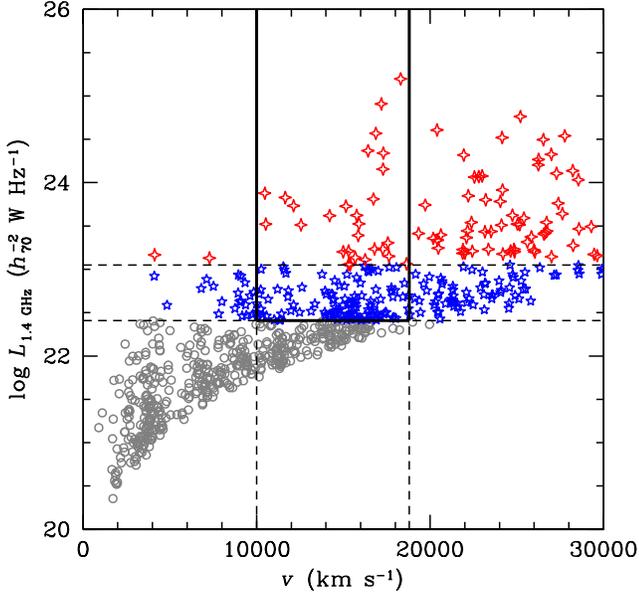}
\vskip -0.1cm
\caption{\label{fig:distrib} Radio luminosity versus radial velocity 
(all radio-detected galaxies with $ v < 30\,000 \kms$ are shown). 
AGN are the \emph{small red open crosses}, SBGs are the \emph{blue stars} and SFGs 
the \emph{gray circles}. \emph{Horizontal lines} delineate the limits for SBGs 
(eq.~[\ref{Lcut}]) and AGN, while the \emph{vertical lines} delineate our region of 
analysis (eq.~[\ref{vcuts}]). The \emph{U-shaped set of solid line segments} delimits 
our subsample of radio-detected galaxies that is doubly complete in volume and radio luminosity. 
} 
\end{figure}

We cross-identify the the 6dFGS+FLASH galaxies with 
the NVSS radio sources
using a conservative search radius of $15''$, 1/3 of the NVSS FWHM
resolution.
We find 810 matches with recession velocities $ v < 30\,000 \kms$. 
To avoid selection effects, we then work with a subsample that is
flux-limited to
\begin{equation}
S_{\rm 1.4\,GHz} \geq S_{\rm min} = 2.8\,\rm mJy \ ,
\label{fluxlimit}
\end{equation}
(for 75\%
  differential completeness) and
volume-limited to a section 
of a wide shell, where the SSC lies at the mean depth:  
\begin{equation}
10\,000 < v < v_{\rm max} = 18\,800 \, \rm km \, s^{-1} \ .
\label{vcuts}
\end{equation}
For a complete selection in radio luminosity with a maximal number 
of radio-detected galaxies selected, we adopt a lower limit to the
radio-luminosity of
\begin{equation}
h_{70}^2\,\log L_{1.4\,\rm GHz} >
h_{70}^2\,\log L_{1.4\,\rm GHz} (S_{\rm min},v_{\rm max}/c) =  22.41 \ ,
\label{Lcut}
\end{equation}
with 
$L_{1.4\,\rm GHz}(S,z) = 4\pi D_L^2(z)\,S\,
(1+z)^{-(\alpha\!+\!1)} \ ,$
where $D_L(z)$ is the cosmological luminosity distance at redshift $z=v/c$,
$c$ is the  
velocity of light, 
and 
the term $(1+z)^{-(\alpha+1)}$ is the product of the $k$-correction for
a power-law radio  
spectrum: $S(\nu) \propto \nu^\alpha$, with $\alpha \simeq -0.7$ (e.g
\citealt{Condon92})  
and the redshift correction for the frequency
normalization.\footnote{Peculiar velocities make the radial velocity an 
  imperfect distance estimator. For clusters with velocity dispersions as
  high as $800 \, \rm km \, s^{-1}$, typical distance errors will be
  $800/14500 = 5\%$ and can be 
 as high as 14\% for peculiar velocities as large as
 2.5 cluster velocity dispersions.
Therefore, the errors
  in relative luminosity will typically be 5\%, and can easily be
  neglected given the several orders of magnitude range in radio luminosities
  (See Fig.~\ref{fig:distrib}).}
In the velocity range of equation~(\ref{vcuts}) lie 2363 galaxies in the merged 
6dFGS+FLASH catalog, which we will refer to as our ``galaxy sample''. 
Among those galaxies, 142 galaxies ($\sim 6$\% of our galaxy sample)
show radio emission above the lower radio flux limit of equation~(\ref{fluxlimit})
and radio luminosity limit of equation~(\ref{Lcut}). 
$85 \%$ of the sources lie within a radius of $10''$.

Following \cite{Sadler02}, we define high radio luminosity galaxies as active galactic 
nuclei (AGN), with lower radio luminosity galaxies as star forming galaxies (SFGs), and 
adopt their minimum AGN radio luminosity of 
\begin{equation}
\log L_{1.4\,\rm GHz} = 23.05 \ ,
\label{AGNcut}
\end{equation}
(after correction to our adopted Hubble constant).
Note that \cite{Machalski00} find that galaxies with AGN spectra predominate starting 
at $\log L_{1.4\,\rm GHz} = 23.31$ (LCRS survey) and 23.05 (UGC galaxies, again converting
both to our adopted Hubble constant). Both \citeauthor{Machalski00} and \citeauthor{Sadler02} 
find AGN with radio luminosities as low as $10^{22} \WHz$
Interestingly, our lower luminosity cutoff at 
$\log L_{\rm 1.4\,GHz} = 22.41$ corresponds to roughly $15 \Msunyr$ --- using
${\rm SFR} ({\rm M_\odot\,yr}^{-1}) = 5.9 \pm 1.8 \times 10^{-22}\, L_{\rm
1.4\,GHz} \, (\rm W\,Hz^{-1})$,
following \cite{Yun01} --- which is close to 
the approximate lower limit for the
definition of a starburst galaxy (SBG).
According to this classification, there are 114 SBGs and 28 AGN 
(respectively $\sim 5\%$ and $1\%$ of our galaxy sample) among the 142 
radio-detected galaxies.
Figure~\ref{fig:distrib} shows our selection as a function of 
radio luminosity and radial velocity (roughly equivalent to distance), 
and highlights our subsample of radio-detected galaxies with the constraints 
on radial velocity (eq.~[\ref{vcuts}]) and radio luminosity (eq.~[\ref{Lcut}]), i.e. 
doubly complete in volume and radio luminosity. 

In addition, we visually checked all identifications 
using radio contours from NVSS, overlaid on optical images from the Digitized 
Sky Survey (DSS). In only a very few cases were the identifications not clear. 
However, we are probably missing a few AGN because the radio emission 
from their lobes could be too far from the optical position of the galaxy to result in a 
match. 
Because our visual checks indicate that gross misidentifications are rare, we expect that 
these more subtle misidentifications should also be infrequent in our final sample, which 
should therefore be valid to perform a statistical analysis.

Throughout the paper, we will consider both radio luminosity and radio loudness 
\footnote{Since we restrict our analysis only to the subsample of radio emitting 
galaxies, the adjective \emph{radio-loud} will refer to 
high ${\cal R}_K$.}.
We use the $K$-band luminosity as a tracer of the galaxy mass, and therefore define 
the dimensionless radio loudness as
\begin{eqnarray}
\log  ({\cal R}_K) &=& \log \left ({S_{1.4\,\rm GHz} \over \hbox{1\,mJy}}
\right ) + 0.4\,K - 5.82 \nonumber \\
&=& \log \left ({L_{1.4\,\rm GHz}\over {\rm 1\,W\,Hz^{-1}}}
\right ) + 0.4\,M_K - 12.90 \ ,
\label{defloudness}
\end{eqnarray}
where we used the normalization of $666.7\,\rm Jy$ for an object with $K=0$ 
\citep{2MASSExpCat}. 
$K$-band magnitudes were retrieved from the 2MASS Extended Source catalog for all
the galaxies in our sample and $K$-band luminosities derived using the FLASH and
6dFGS redshifts. 
We computed $k$ corrections with the approximation $k_K = -1.688 \, z + 3.458 \, z^2$, 
which appears to be a good fit to better than 0.01 mag for $z \!<\! 0.1$ to the tabulated 
Sa galaxy $k$-corrections of \citet{Poggianti97}.
The FLASH and 6dFGS surveys also provide $b_J$ magnitudes, measured with
  the APM and SuperCosmos machines, respectively. We 
$k$-corrected them with $k_{b_J} = 4.04\,z+2.00\,z^2$, again a fit (better
  than 0.001 mag for $z < 0.1$) to
  \citeauthor{Poggianti97}'s $k$-corrections for an Sa galaxy in the nearby $B$ band. 

Because $K$-band luminosity is closely (but not perfectly) related to stellar mass, 
${\cal R}_K$ can be thought a measure of stellar mass-weighted efficiency of
instantaneous star formation for SBGs. For AGN, which usually have higher  
radio to optical ratios, it gives an idea of the radio power of the nuclei 
in comparison to the stellar mass of the galaxy.

\subsection{Large scale structure}
\label{subsec:lss}

\begin{figure*}
\begin{center}
\resizebox{\hsize}{!}{\rotatebox{90}{\includegraphics[bb=150 186 480 706]{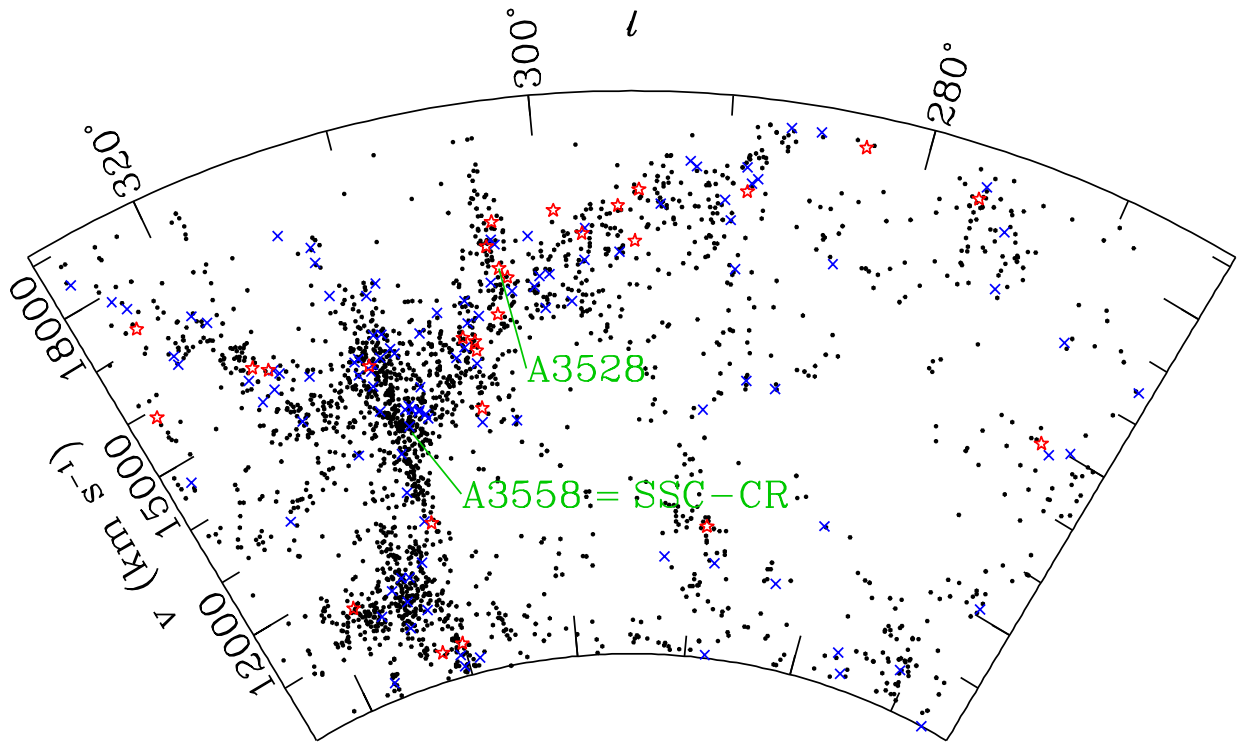}}}
\end{center}
\vskip -0.5cm
\caption{\label{fig:wedges} Wedge plot in galactic longitude and
    heliocentric velocity of the FLASH 
survey area showing both FLASH and 6dFGS
galaxies (\emph{black dots}) and cross-identified
radio-detected galaxies: AGN are shown as \emph{red open stars} 
and SBGs as \emph{open 
blue crosses}. 
The SSC is the Y-shaped large-scale structure in the left part of the
  plot. The two cluster complexes studied here are shown.
}
\end{figure*}

The wedge plot of Figure \ref{fig:wedges} shows the three dimensional 
distribution of galaxies in our 6dFGS+FLASH catalog, projected in
galactic longitude.
 This region, which
spans $36\,\rm Mpc$ by $251\,\rm Mpc$ in galactic coordinates, is filled 
with 19 Abell clusters of galaxies in the velocity range of
equation~(\ref{vcuts}). 
The SSC is the large overdensity centered on A3558
$(13^h27^{m}, -31^{\circ}29^{'}, v=14\,390 \kms)$, with the filaments
  extending away.

In this paper, we define the SSC-CR (or the A3558 cluster complex) as 
a $10 \Mpc$ radius cone, limited to velocities within $3\,\sigma_v$ of the
  set of 
  clusters A3556, A3558, A3559, A3560 and A3562 that compose it
  ($11459 \, \rm km \, s^{-1} < v < 17545 \, \rm km \, s^{-1}$).
The complex has a dense core with
clusters A3562, A3558, A3556 and groups SC 1329-312, SC 1327-313, with
  clusters A3559 and A3560 in its envelope.
Altogether, we have 423 galaxies in the SSC-CR, among which 20 SBGs, but not
  a single AGN.
Another prominent feature to the West of the supercluster 
is the A3528 cluster complex, centered on A3528 at 
$(12^h54^m,-29^{\circ}01', v=15\,829 \kms )$. The A3528 complex is 
defined in a similar manner as the SSC-CR: a cone of radius $10\,\rm Mpc$, 
with velocities within $3\,\sigma_v^{\rm A3528}$: $13069 \, \rm km \,
  s^{-1} < v < 18771 \, \rm km \, s^{-1}$. It
  thus contains 
clusters 
A3528, A3530 and A3532.
We have 121 confirmed galaxies in this complex, among which 5 SBGs and 5 AGN.
Other features visible here are described in greater detail in
  \cite{Quin95,Quin00}.  

\begin{figure}
\hspace{5cm}
\centering
\includegraphics[width=8.5cm]{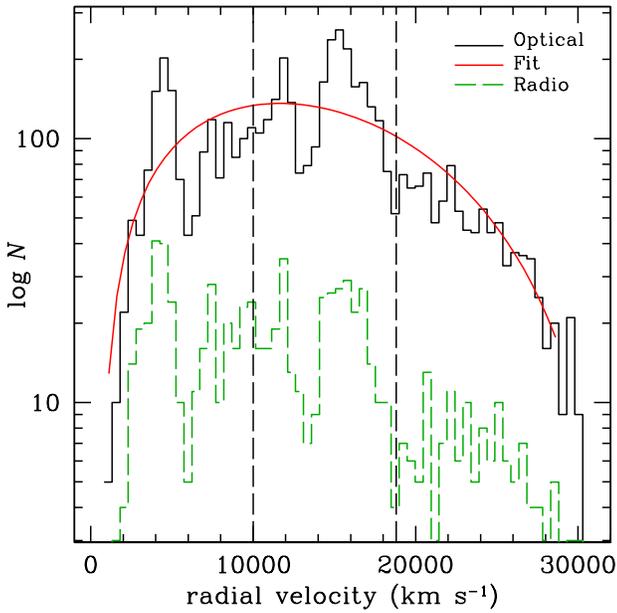}
\vskip -0.1cm
\caption{\label{fig:numgals} Distribution of radial velocities of the optical/NIR (\emph{solid 
black histogram}) and radio (\emph{green dashed histogram}) samples. The two \emph{dashed 
vertical lines} delineate our velocity cuts (corresponding to the rough velocity limits of the SSC). 
The \emph{red curve} is the 4th order polynomial fitted to the underlying optical/NIR distribution 
and represent the empirical selection function of the data.}
\end{figure}

Figure~\ref{fig:numgals} shows the redshift distribution for the 6dFGS+FLASH 
galaxy catalog and the radio galaxy subsample. 
The two outstanding peaks in the observed redshift distribution are the Hydra cluster 
at $\sim 4000 \kms $ and the SSC at $\sim 14\,400 \kms$.
The large transverse filament at $16000 \kms$ seen in the wedge plots of Figure~\ref{fig:wedges} 
also produces a wide peak in the redshift distribution histogram. 

Both figures~\ref{fig:wedges} and \ref{fig:numgals} show that, at first glance, 
\emph{radio-detected galaxies trace fairly well the underlying galaxy distribution}:  
radio-detected galaxies are mostly found in the dense regions and less likely in the voids. 
However, Figure~\ref{fig:wedges} indicates that 
the SSC-CR seems is devoid of strong radio sources, which
cluster  
more in the A3528 cluster complex and its surroundings. 

Since our radio galaxy classification is based upon 
a luminosity cut, it therefore creates a strong distance segregation, 
caused by Malmquist bias, where only luminous radio 
galaxies are present at larger distances. 
But since we restrict our analysis to a subsample doubly complete in volume
and radio luminosity, Malmquist bias is no longer a concern.

\section{Radio emission in the A3558 and A3528 complexes}
\label{sec:radiofracs}

According to our definition of the cluster complexes
  (Sect.~\ref{subsec:lss}), 
there are 1819 galaxies, among which 88 SBGs and 23 AGN, in the remaining 
FLASH+6dFGS area (outside both cluster complexes and 
with the velocity cut of eq.~[\ref{vcuts}]).
This sample will be used as a comparison sample 
and hereafter referred to as our \emph{reference sample}.

The fraction of radio-detected galaxies (SBG+AGN)
are 4.7\%, 8.3\% and 6.1\% in the A3558 complex, the A3528 complex,
and the reference sample.
Binomial statistics indicate that the lower (higher) fraction of radio
galaxies ($\log L_{1.4\,\rm GHz} \geq 22.41$) in the A3558 (A3528)
complex relative to the reference sample is not significant.

Considering separately SBGs and AGN (with eq.~[\ref{AGNcut}]),
we find 20 SBGs and no AGN in the A3558 cluster complex, 
whereas there are 5 AGN and 5 SBGs in the A3528 cluster complex.
Binomial statistics indicate that the absence of AGN among 20 
radio-detected galaxies 
in the A3558 complex would occur by chance only 1.0\% of the time, given the observed 
fraction ($23/111 = 21\%$) of AGN in the reference sample.
On the contrary, the A3528 complex seems to harbor too many AGN relative to the reference 
sample, as there is only a 3.8\% chance of obtaining at least 5 AGN among 10 radio 
galaxies given the fraction of radio-detected galaxies that are AGN in the reference sample. 

\begin{figure}
\hspace{5cm}
\centering
\includegraphics[width=8.5cm]{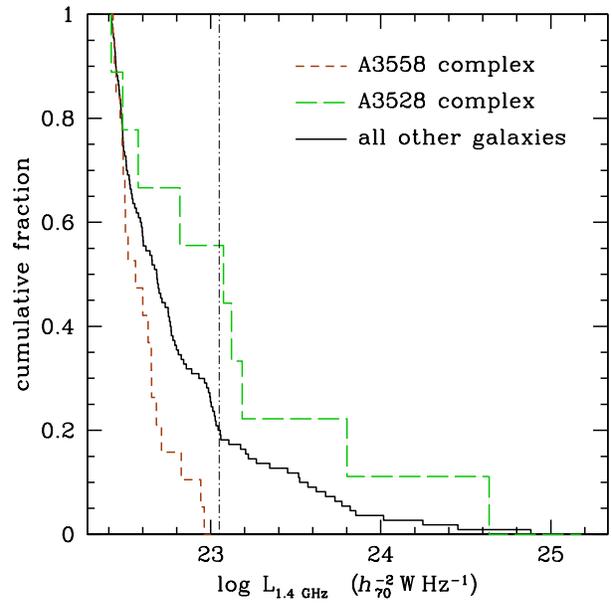}
\includegraphics[width=8.5cm]{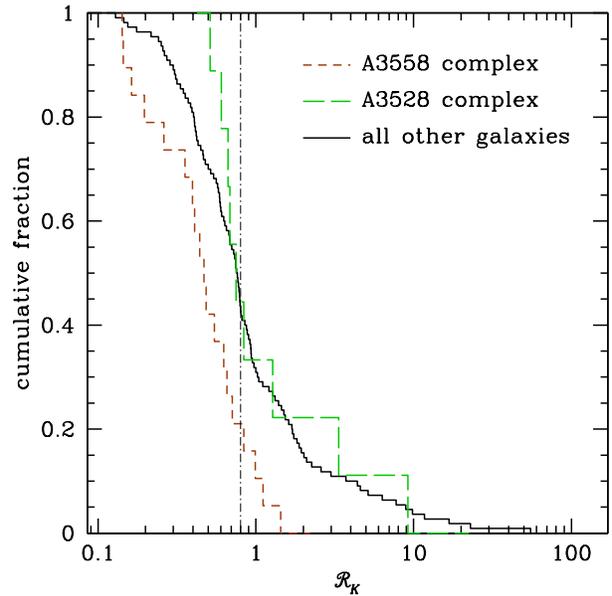}
\vskip 0.2cm
\caption{\label{fig:ks} Normalized cumulative distributions of radio luminosity 
(\emph{top}) and radio loudness (\emph{bottom}) for three regions: the A3558 complex 
(\emph{brown short-dash histogram}), the A3528 complex (\emph{green long-dash histogram}) 
and the reference sample (\emph{black solid histogram}).
The \emph{vertical lines} denote the transition from SBGs to
AGN (\emph{top plot}) and our ad hoc separation of radio loudness
(\emph{bottom plot}).
}
\end{figure}

The top plot of Figure~\ref{fig:ks} shows the cumulative radio
  luminosity function (RLF) of the
  A3558 and A3528 complexes.
Of course, one can argue that it is dangerous to compare luminosities of samples at 
different distances, even with the same magnitude limit, as the more distant sample 
will lead to more luminous galaxies.
However, A3528 is only 10\% more distant than A3558, which should lead to luminosities 
typically 20\% larger, i.e. a 0.08 increase in $\log L_{1.4\,\rm GHz}$. 

The plot clearly indicates a lack of high radio luminosity galaxies in the
A3558 complex, which is marginally significant (94\% confidence with a
Kolmogorov-Smirnov --- KS ---
  test, 95\% with a Wilcoxon rank sum test).

However, for low radio luminosities, the radio galaxy distribution in 
the A3558 cluster complex resembles that in the reference sample, while the 
two distributions depart from one another only at intermediate and high 
radio luminosity.
On the other hand, the A3528 complex shows an RLF that appears shifted to higher
radio luminosities in comparison with the RLF of the reference field, but the
difference is only marginally significant (91\% confidence with a KS test).

Splitting radio-detected galaxies into two classes, above and below an arbitrary limit of 
${\cal R}_K = 0.8$ (the median is 0.7), 
we find only 4 radio-loud galaxies out of 20 in the A3558 cluster 
complex (20\%) and 5 in the A3528 complex (50\%).
In the reference sample, the fraction of radio-loud galaxies is 0.43. 
Binomial statistics indicate that the probability that the A3558 
complex would have as few as 4 galaxies with ${\cal R}_K > 0.8$ amongst a total of 20
radio-detected galaxies is 3\%: the lack of radio-loud galaxies 
in the A3558 complex is statistically significant. This conclusion depends
little on our cut in radio loudness in the range $0.6 < {\cal R}_K < 0.9$.

The bottom plot of Figure~\ref{fig:ks} illustrates 
this lack of radio-loud galaxies in the A3558 complex:
\emph{the distribution of ${\cal R}_K$ in the A3558 complex 
is shifted to values $\approx 2\times$ lower than for the reference
  field}, 
and
a KS test gives a 4\% probability of a greater difference by chance
(only 1\% with a Wilcoxon rank sum test).
On the other hand, the distribution of radio-loudness in the A3528 complex is
shifted to higher values, but because of small number statistics this offset
 is not significant.

To summarize, the two cluster complexes show different trends in their
distribution of radio  
luminosities and radio loudness relative to the reference sample. 
\emph{the A3528 complex 
galaxies are marginally more radio-luminous than the galaxies in
  the reference sample, while the
A3558 complex galaxies are marginally less radio-luminous and significantly
less radio-loud.}

These distributions tell us that the special dynamical state of the 
very dense core of the Shapley supercluster has a quantifiable
impact on its radio galaxy population that significantly reduces the radio activity.
However, these radio luminosity and radio-loudness \emph{distributions} 
do not allow us to quantify what is the impact of their respective
\emph{environments}. 
We therefore investigate in the next section how these variables are 
affected by galaxy density, at the supercluster scale as well as inside clusters within 
the cluster complexes to determine what environmental processes might be at work.

\section{Modulation of radio emission with the density of the environment}
\label{sec:envmt}

\subsection{Estimation of the density of the environment of radio-detected galaxies}
\label{subsec:smoothing}

As seen in Figure~\ref{fig:numgals}, 
the fraction of radio-detected galaxies is independent of
redshift. This means that 
we can estimate the density of the environment using the
optical sample.

Given the discrete nature of the galaxy positions, we choose to measure a continuous 
density. We thus define a continuous density in redshift space, by smoothing the discrete 
density in redshift space with a Gaussian kernel of scale $\sigma_{rz}$ 
(c.f. \citealp{Tully88,Monaco94}), using:
\begin{equation}
\rho_{rz,j}\, \left [h^3 {\rm Mpc^{-3}}\right] = 
\frac{1}{\left (2 \pi \sigma_{rz}^2 \right )^{3/2}} 
\sum_{i \neq j}  
\exp \left ( \frac{-r_{ij}^2}{2 \,\sigma_{rz}^2} \right )  \ ,
\label{rhorz}
\end{equation}
where subscripts $j$ and $i$ respectively represent radio and optical
galaxies and where 
\begin{eqnarray}
r_{ij} &=& \sqrt{d_p^2(z_1)+d_p^2(z_2)-2\,d_p(z_1)\,d_p(z_2)\,\cos
  \theta_{ij}}
\nonumber \\
&\simeq& {c\over H_0}\,\sqrt{z_{ij}^2+(\theta_{ij}\,\langle z\rangle)^2} \ ,
\label{rij}
\end{eqnarray} 
is the separation of two galaxies in redshift space\footnote{We actually use
  the cosmological formula, only valid for a flat Universe 
(first equality of eq.~[\ref{rij}]), 
but at the small 
redshift of our sample ($z \!<\! 0.06$), cosmological corrections are typically
  less than 4\%.}, 
with $\theta_{ij}$ their angular separation and $\langle v \rangle$ their 
mean radial velocity, and where $d_p(z)$ is the proper distance.

Of course, such an estimator will be subject to edge effects (the smoothed density
will decrease near the survey edges) and to the radial selection function inside the
area. To correct for this, we generated 25 random datasets, each with as many
galaxies as in our observed sample, in the exact same survey geometry and with the 
same velocity selection function, approximated with a 4th order polynomial fit to 
the velocity distribution, as shown in Figure~\ref{fig:numgals}.  
Note that the mean local densities of the 25 random catalogs obtained with 
equation~(\ref{rhorz}) will be 1/25th of the local densities of a single random 
catalog with 25 times as many galaxies. For each smoothing scale, we calculate the 
Gaussian-smoothed density and divide it by the mean Gaussian-smoothed density of 
our random catalogs.

In redshift space, the 3D density is biased by redshift distortions, i.e. incorrect 
radial positions of cluster galaxies due to fingers of God. We therefore also
derive a continuous surface  
density of galaxies by Gaussian smoothing the discrete surface density, as:
\begin{equation}
\rho_{\theta,j}\,\left [h^2 {\rm Mpc^{-2}}\right] = \frac{1}{2 \pi
\sigma_\theta^2 } \sum_{i \neq j}  
\exp \left ( \frac{-\theta_{ij}^2}{2 \sigma_{\theta}^2} \right )  \ ,
\label{rhotheta}
\end{equation}
again dividing by the mean surface density of the 25 random catalogs.
Even though both approaches suffer biases, the combination of the two will
allow us to check for consistencies in the trends.

We calculated densities for a wide range of smoothing scales, each separated 
by a factor of 2: $\sigma = 0.625,\,1.25,\,2.5,\,5,\,10,\,$ and $20 \,
h_{70}^{-1} \, \rm Mpc$.\footnote{For the density in projected space, we 
used for $\sigma_\theta$ in equation~(\ref{rhotheta}) angles of 
$\sigma/(200\,\rm Mpc)$, corresponding to the angular sizes of the redshift space 
smoothing scales at the distance of the SSC. In what follows, we will replace 
$\sigma_\theta$ by the equivalent scale in Mpc.}  
Of course, large scales suffer from a smaller range of smoothed densities.
Small scales will suffer from the lack of close neighbors (in both the real
and random catalogs).
In addition, the surveys are not complete in crowded areas because of fiber avoidance 
and we therefore underestimate the density in the denser regions.

We illustrate our procedure by showing, in the top panels of
Figure~\ref{fig:smooth}, the modulation of near-infrared luminosities with
density, for two smoothing scales. 
One can easily notice a trend of luminosity segregation: the more
luminous galaxies prefer the dense environments, especially at a smoothing
scale of 2.5 Mpc, which is roughly the scale of clusters.
Spearman tests indicate a rank correlation of $-0.32$ and $-0.22$ at 2.5 and
10 Mpc, respectively, which are over 99.5\% significant.
\emph{At a smoothing scale of 1.25 Mpc, $K$-band luminosity is significantly
  anti-correlated with density, regardless of whether density is measured in
  projected or angular space.} 
But at smoothing scales $\geq 2.5$ Mpc, $L_K$
is not significantly anti-correlated with projected density.

\subsection{Radio luminosity -- density relation} 
\label{subsec:densradio}

\begin{figure}
\centering
\includegraphics[width=\hsize,bbllx=50,bblly=0,bburx=594,bbury=762]{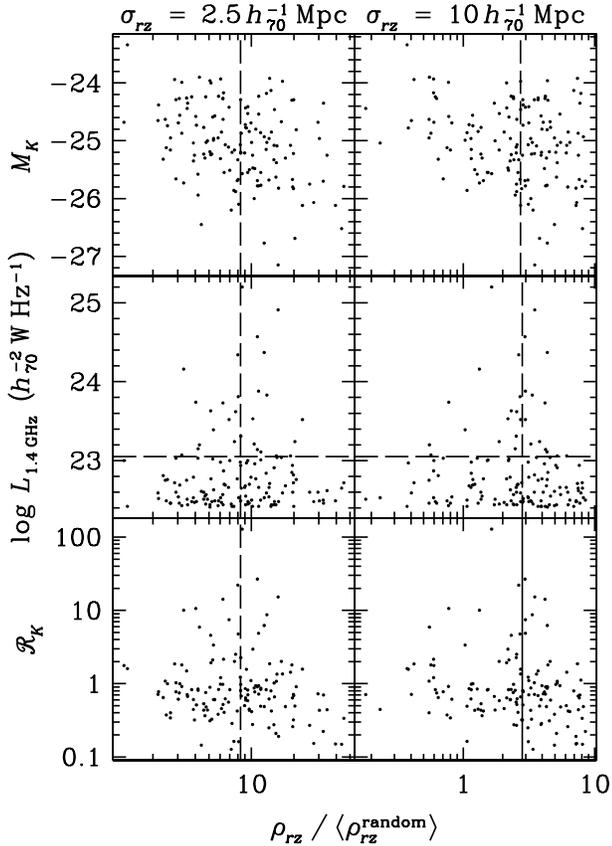}
\caption{\label{fig:smooth} Gaussian-smoothed density in redshift-space 
corrected for edge effects and the radial selection function 
versus $K$-band luminosity
(\emph{top plot}), radio luminosity 
(\emph{middle plot}) and radio loudness 
(\emph{bottom
  plot}), for two different smoothing scales.
The \emph{dashed vertical} and \emph{dashed 
horizontal lines} represent the limit between high and low density environment 
and the separation between SBGs and AGN, respectively (see text).}
\end{figure}

The middle panels of  Figure~\ref{fig:smooth} 
show that
the largest radio luminosities appear to prefer intermediate densities, near
the median. 
However, the distribution of radio luminosities of galaxies at
 intermediate densities is
 not significantly different from the analogous distribution at extreme (high
 and low) densities. And, for all smoothing scales, 
there are no significant correlations of radio
 luminosity with both redshift space density and projected density.

\subsection{Radio loudness -- density relation}
\label{subsec:densradioK}

The correlation between a galaxy radio luminosity and the density of its environment 
might hide a correlation between radio luminosity and mass (as both are extensive variables) 
on one hand, and mass and environmental density on the other hand (as expected in models 
of galaxy formation where the more massive galaxies are more clustered). 
In other words, 
since more luminous galaxies are usually more clustered, and should have undergone 
several major mergers, they might also be prone to radio activity. We
therefore 
 correlate the radio 
\emph{loudness}
(eq.~[\ref{defloudness}]) with the density of the galaxy environment.

As seen in the bottom right panel of Figure~\ref{fig:smooth},
for smoothing scale $10 \, \Mpc$,
radio loudness appears to be anti-correlated
 with the density of the environment:
the low loudness galaxies are virtually all at high densities.
This is quantified in 
Figure~\ref{fig:cutoutL},
which indicates \emph{highly significant anti-correlations of radio loudness 
with the density of the environment, measured on scales greater than 
$2 \,  \rm Mpc$}. These significant anti-correlations occur in both projected 
and redshift spaces, which gives us confidence in these observed trends. 
For example, at a smoothing scale $\sigma = 10 \,  \rm Mpc$, 
both $\rho_{\theta}$ and $\rho_{rz}$ are negatively correlated ($r \sim
-0.17$ and $r \sim -0.21$) with ${\cal R}_K$ with a high
significance (98\% and 99.4\%, respectively). 

We also used Kolmogorov-Smirnov (KS) tests to compare
the distributions of densities for AGN vs. SBGs, as well as the distributions of radio
luminosities, loudness (and NIR luminosities) for galaxies in high vs. low
density environments. No significant 
trends were found at any smoothing scale
in both analyses of projected and redshift space densities.
This shows that the KS test
is not as an efficient estimator of the environmental effects  as
is the rank correlation with density.

\subsection{Cutting out the cluster complexes}
\label{subsec:removSSC}

\begin{figure}
\centering
\includegraphics[width=8.5cm]{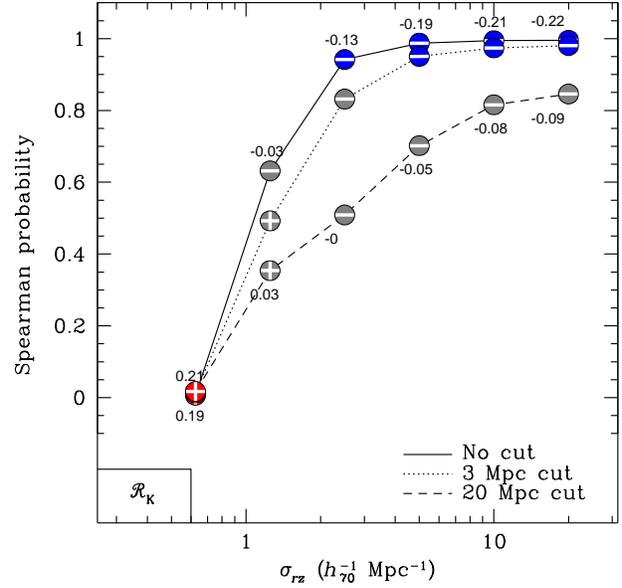}
\caption{\label{fig:cutoutL} Probability of statistical significance of Spearman 
rank correlation of 
radio loudness 
with Gaussian-smoothed density in redshift space (normalized with the mean
density derived from the random catalogs)
versus the smoothing scale.
\emph{Dotted} and \emph{dashed curves} represent the analysis with the
central portion of the SSC cut out to a radius of respectively
$3 \, \rm Mpc$ and $20 \, \rm Mpc$, and limited to
  $\pm 3\,\sigma_v \hbox{(A3558)}$.
\emph{Red positive signs} highlight positive correlation coefficient 
and \emph{blue negative signs} 
the negative ones. The Spearman probability is defined as 
$P_{\rm Spearman} = (P_{\rm chance} - 1/2)  \, {\rm sgn}(r) + 1/2$.
The values of the rank correlation coefficient $r$ are given for the two
extreme cases.}
\end{figure}

To better understand the anti-correlation of radio
 loudness with density, 
we performed the same analysis as in Sect~\ref{subsec:densradioK} on the
 FLASH area, but with parts of the SSC cut out 
from the sample. 
We calculate densities only for the radio-detected galaxies 
outside the A3558 cluster complex, whose projected radius is now a free
  parameter.
Figure \ref{fig:cutoutL} shows how the significance of correlation ($P \to 0$) or
anti-correlation ($P\to 1$) of radio loudness with redshift space density varies with the
smoothing scale, for different choices of the projected radius of the SSC-CR
we cut from our sample.

At large cut sizes,
\emph{the anti-correlation between radio loudness and local density at 
large scales disappears when removing the central region of the Shapley supercluster} 
from the statistical analysis. 
The change in the statistical trend 
becomes visible at a cut size of $ \sim 4 \, \rm Mpc$. As the projected radius
of the removed area 
increases, the anti-correlation vanishes. When reaching a cut size of $ \sim 6 \, \rm Mpc$, 
all the Spearman probabilities drop below $90\%$. 
However, if we perform a similar analysis by removing a $20\, \rm Mpc$ shell
centered on A3528, we do not see any loss of anti-correlation at large scales, 
which means that this area is not responsible for the anti-correlation. 

Since the SSC-CR seems to be the main area that
disfavors the presence of radio-loud galaxies, we now 
compare the A3558 complex with its surroundings.
As can be seen in 
Figure~\ref{fig:corrsel},
there is a strong anti-correlation between radio loudness 
and the $\sigma_{rz} = 10 \,\rm Mpc$ smoothed density among just the SSC-CR members.
The Spearman rank-correlation coefficient is $r = -0.41$, which indicates a
97\% level confidence for the negative trend. 
Outside the SSC-CR, radio loudness is uncorrelated with density.
As noted before, there are virtually no radio-loud
galaxies in the SSC-CR. Moreover, there seems to be an excess of galaxies
of low radio loudness
in the densest environments at $\sigma_{rz} = 10 \,\rm Mpc$.

\section{Modulation with the relative clustercentric distance}
\label{sec:cluster}

We also identify radio-detected galaxies within and outside of clusters in the SSC-CR. For
each radio galaxy, we search clusters whose redshift matches the galaxy to
within 3 cluster velocity dispersions (measured with NED, which although
  inhomogeneous is more complete than 6dFGS).
The smallest projected
clustercentric distance, normalized to the cluster \emph{virial radius},
  $r_{200}$,  where the mean total mass density of the cluster 
  equals 200 times the critical density of the Universe,
determines which cluster the radio galaxy is closest to.  
This avoids assigning more than one cluster to each radio
galaxy.
We estimated $r_{200}$ from the cluster velocity dispersions according
  to $r_{200} = \sigma_v / 436 \, \rm km \, s^{-1}$, appropriate for the 
NFW model of concentration $c=5$, as
derived in equation~(\ref{rvirsigv}).
We then separate the galaxies within the cluster inner regions, defined as
$0.8\,r_{200}$, from the galaxies
in the outer regions (extending to far 
beyond the cluster virial radius).

\begin{figure}
\centering
\includegraphics[width=\hsize]{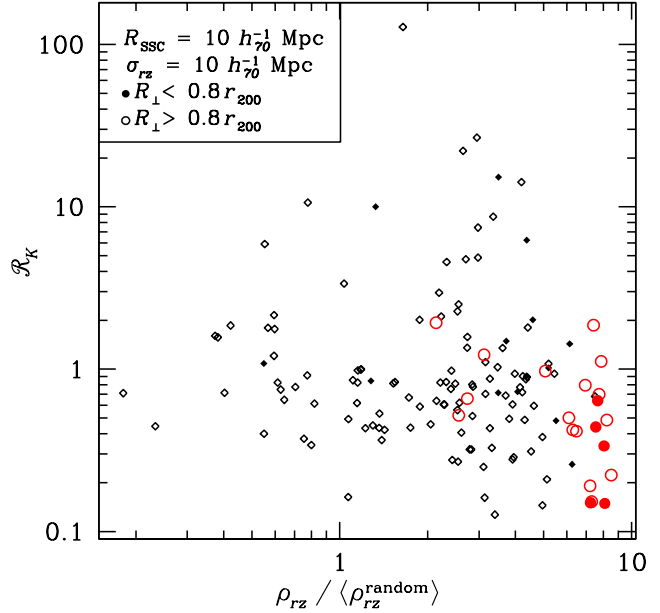}
\caption{\label{fig:corrsel} Radio loudness ($K$-band normalized)
  vs. density (smoothed on scale $\sigma=10 \, \rm Mpc$),
  separating the regions  
within (\emph{big red circles}) and outside (\emph{small black
  diamonds}) the SSC-CR. 
\emph{Filled symbols} show radio-detected galaxies that are less than
  $0.8\,r_{200}$ of a galaxy  
cluster center whereas \emph{open symbols} are the ones located at more
  than $0.8\,r_{200}$.} 
\end{figure}
Figure~\ref{fig:corrsel} shows that 
outside of clusters, the distribution of radio loudness values is the
same for regions within and outside the SSC-CR.
However, Figure~\ref{fig:corrsel} also indicates that,
\emph{within the SSC-CR, the cluster radio-detected galaxies have lower
(typically by a factor 3)  radio loudness than galaxies in clusters outside
the SSC-CR.}

Table~\ref{tab:KSinout} provides KS test probabilities that the differences in
the distributions of various parameters within and outside the SSC-CR are
caused by chance, for
radio-detected galaxies outside of clusters (first line) or for those within clusters
(second and third lines, the latter omitting the central brightest cluster galaxy,
hereafter BCG).
The KS test indicates that the 
3 times lower values of ${\cal R}_K$ of the cluster 
radio-detected galaxies in the SSC-CR relative to their counterparts outside of the 
SSC-CR is very highly significant: there is a 99.7\% probability that the 
difference in distributions of radio loudness is not caused by chance. 
\begin{table}[ht]
\begin{center}
\caption{Significance (with KS test)
of same distributions within and outside of the SSC-CR
\label{tab:KSinout}
}
\begin{tabular}{r|llll}
\hline
\hline
\multicolumn{1}{c|}{Galaxy position}& \multicolumn{1}{c}{$L_{1.4\,\rm GHz}$} &
\multicolumn{1}{c}{$\ {\cal R}_K$} &
\multicolumn{1}{c}{$M_K$} & \multicolumn{1}{c}{$B-K$}\\
\hline
$R>0.8\,r_{200}$ & \ 0.14 & 0.58  & 0.55 & \ 0.95\\
$R<0.8\,r_{200}$ & \ 0.10 & 0.003 & 0.05 & \ 0.77\\
$0.04\,r_{200} < R<0.8\,r_{200}$ & \ 0.32 & 0.01 & 0.08 & \ 0.97\\
\hline
\end{tabular}
\end{center}
\noindent \emph{Note}:
The galaxy position is relative to the center of the nearest cluster.
\end{table}

Now, one could argue that the depressed loudness of radio-detected galaxies
within 
clusters inside the SSC-CR may signify an intrinsic 
anti-correlation between radio loudness and NIR luminosity coupled with a 
correlation between NIR luminosity and density (luminosity
  segregation).  
However, as one can see in Table~\ref{tab:KSinout}, while the differences in
the distributions 
of $K$-band luminosities between cluster radio-detected galaxies within 
and outside of the SSC-CR are  significant,
they are not as significant as the differences in
the distribution of radio-loudness values, even if the  BCG is removed.

Since there are no highly radio-luminous galaxies in the SSC-CR, which we
interpret as an absence of AGN, the depressed values of ${\cal R}_K$ for the
radio-detected galaxies within SSC-CR clusters may be an effect of
enhanced morphological segregation within Shapley clusters: an enhancement of
the early-to-late ratio among spiral galaxies would lead to reduced star formation
efficiencies for spiral galaxies within clusters of the SSC.
And yet, as seen in Table~\ref{tab:KSinout}, the colors of radio-detected galaxies
lying in clusters are not affected by their presence within the SSC-CR.
This suggests that morphological segregation is not the cause of the
significantly lower values of radio loudness for cluster galaxies in the
SSC-CR relative to the cluster galaxies outside the supercluster core. 

Hence, \emph{the physical parameter of radio-galaxies within clusters that 
is most affected by the SSC-CR environment is radio loudness.}
On the other hand, the distribution of radio-luminosity and radio loudness 
in the outer regions of clusters is unaffected by their presence within or out
of the SSC-CR.
Therefore, \emph{the anti-correlation of radio loudness with density of the
  environment is caused by the decreased radio loudness of radio-galaxies
  within SSC-CR clusters, relative to their counterparts outside of the
  SSC-CR.}

\begin{figure}[ht]
\centering
\includegraphics[width=\hsize]{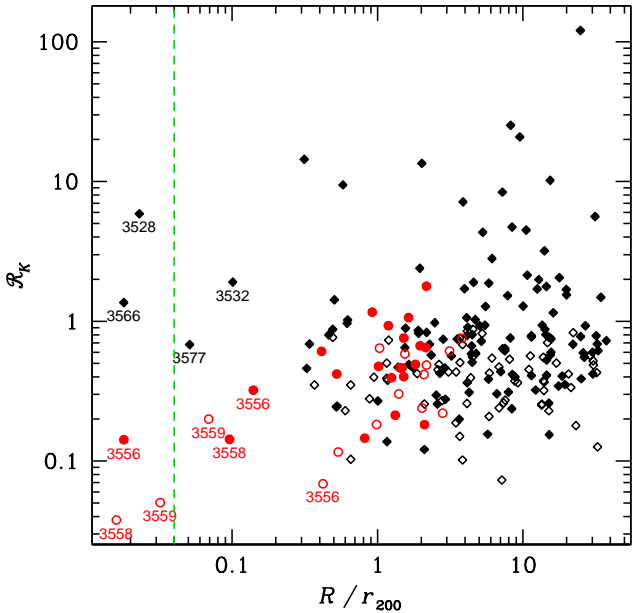}
\caption{Radio loudness ($K$-band normalized) vs. projected distance to
    nearest cluster (in units of cluster virial radius).
\emph{Red circles} and \emph{black diamonds} represent the galaxies
    within and outside the SSC-CR, respectively.
The \emph{filled} and \emph{open circles} represent the galaxies with
    $\log L_{\rm 1.4\,GHz} \geq 22.41$ (complete radio luminosity sample in
    entire region)
and $22.21 \leq \log L_{\rm 1.4\,GHz} < 22.41$ (extension of sample, complete 
    for the A3558 complex).
The labels indicate the Abell cluster number of the galaxies with the lowest
    values of radio loudness.
The galaxies \emph{left of the vertical line} are brightest cluster
    galaxies, actually lying at the cluster center.
\label{fig:loudvspos}}
\end{figure}

The suppression of radio emission in the SSC cluster galaxies is even clearer
in Figure~\ref{fig:loudvspos}.
As seen in Table~\ref{spearx200}, inside the SSC-CR, 
radio loudness and relative position in the nearest cluster are indeed
significantly correlated ($r=0.50$, 99\% confidence), while outside the
SSC-CR there is no correlation between radio loudness and clustercentric
position. 
Part of the loudness - radius correlation in the SSC-CR 
may be due to suppressed radio
emission in the BCGs sitting in the cluster
cores \citep{Burns90}.
Since we used the Abell cluster centers 
given by \cite{ACO89}, which do
not coincide with the positions of the BCGs, these central galaxies are visible
in Figure~\ref{fig:loudvspos}. We visually checked
with NED the positions of the BCGs and found all of them to lie within
$0.04\,r_{200}$.

The correlation between radio loudness and
clustercentric radius remains significant ($r=0.42$, 96\% confidence)
once the BCGs ($R/r_{200} < 0.04$, given the imprecise centers that we have
used)
are removed.
Figure~\ref{fig:loudvspos} also shows an extension of the sample to lower
radio luminosities, which is complete for the SSC-CR, but not for the full
region outside the complex. The trends for low radio loudness of BCGs in the
SSC-CR are confirmed, as is the same (weaker) trend for non BCGs within
$0.3\,r_{200}$ of the cluster center.

\begin{table}[ht]
\begin{flushleft}
\begin{center}
\tabcolsep 4pt
\caption{Rank correlations with clustercentric radius and their significance
  (Spearman test)
\label{spearx200}
}
\begin{tabular}{lc|cccc}
\hline
\hline
Zone & \multicolumn{1}{c|}{$R/r_{200}$}& \multicolumn{1}{c}{$L_{1.4\,\rm GHz}$}
& 
\multicolumn{1}{c}{$\ {\cal R}_K$} & 
\multicolumn{1}{c}{$M_K$} & \multicolumn{1}{c}{$B-K$}\\
\hline
 in &  all       & --0.38 \ (95) & \ \  0.50 \ \ \ (1) & \ \  0.75 \ \ \ (0) & --0.27 \ (88)\\ 
 in &  $>0.4$    & --0.35 \ (93) &  \ \ 0.42 \ \ \ (4) & \ \  0.70 \ \ \ (0) & --0.19 \ (79)\\ 
\hline
out &  all       & --0.08 \ (81) &  \ \ 0.01 \ (46) &  \ \ 0.17 \ \ \ (3) &  \ \ 0.04 \ (32)\\ 
out &  $>0.4$    & --0.05 \ (69) &  \ \ 0.04 \ (32) &  \ \ 0.15 \ \ \ (5) &  \ \ 0.07 \ (22)\\ 
\hline
\end{tabular}
\end{center}
\end{flushleft}
\noindent{\footnotesize 
{\em Notes}: Column (1): within or outside SSC-CR; column (2): range of projected
distances to the cluster center
normalized to $r_{200}$; following columns: Spearman rank correlation coefficient, $r$,
and associated probability $P_{\rm Spearman} = (P_{\rm chance} -
1/2)  \, {\rm sgn}(r) + 1/2$ in percent in parentheses. 
The correlations all involve samples with $\log
L_{\rm 1.4\,GHz} \geq 22.41$, so that all samples are complete in radio luminosity.}
\end{table}

Table~\ref{spearx200} also shows that in the SSC-CR, 
radio luminosity is marginally
correlated with clustercentric position, but this correlation is absent
outside the SSC-CR.
On the other hand, \emph{the NIR luminosity is very
strongly correlated with clustercentric radius, especially so 
in the SSC-CR, but also outside the
cluster complex
(i.e., NIR absolute magnitude is very strongly anti-correlated with
clustercentric radius). 
The respectively strong and weak anti-correlations of NIR and radio luminosity
with clustercentric radius inside the SSC-CR
explain the positive correlation of radio loudness
with clustercentric radius within the SSC-CR.}

One could worry that the radio loudness - clustercentric radius
  correlation in the SSC-CR could be the consequence of the combination of 1)
  the strong NIR luminosity segregation, 2) the weak radio luminosity
  segregation, and 3) the radio luminosity limit of our sample.
But since our sample is strictly limited in radio luminosity and virtually so
  in NIR luminosity (given the $K$-band limit of the 6dFGS sample and the
  volume limit that we imposed), selection effects should not affect radio
  luminosity significantly more than they affect NIR luminosity. Therefore,
  it is difficult to understand how selection effects could lead to a
  significant correlation of radio loudness with clustercentric radius in the
  SSC-CR.
It therefore appears that \emph{the suppression of radio loudness
  is independent of the strong segregation in NIR luminosity, even when
  removing BCGs from our sample}.

\section{Comparison with previous studies}
\label{sec:compare}

The distributions of radio luminosities in the A3558 and A3528 cluster
  complexes have  
previously been studied by \cite{Venturi00}, \cite{Miller05} and \cite{Venturi01}.
\cite{Venturi00} and \cite{Miller05} 
found that the cumulative radio luminosity function (RLF)
of
galaxies in the A3558 complex has a steeper bright-end slope than the
  respective reference samples of \cite{Ledlow96} (E/S0 galaxies only) and \cite{MO01}.
This is in close qualitative agreement with the RLFs of the A3558 complex and the
  reference field that we have shown in the top plot of
  Figure~\ref{fig:ks}. However, 
the radio luminosity where the SSC-CR RLF begins to depart from the reference
one
occurs at 
  slightly different radio luminosities: $\log L_{1.4\,\rm GHz} \la 22.31$
  for \citeauthor{Venturi00} (corrected to our cosmology), 
compared to $\log L_{1.4\,\rm GHz} = 22.6$ for
  \citeauthor{Miller05} and 22.45 for us.

\citeauthor{Venturi00} also noted that the optically-defined radio loudness
was uncorrelated with 
the density of the environment.
Similarly, \citeauthor{Miller05} did not see any spatial trends of
optically-defined radio loudness in the SSC-CR. 
Both results are in sharp contrast with the strong anti-correlation of NIR-defined
  radio loudness with the density of the environment and the strong
  correlation of radio loudness with clustercentric radius that we found
for radio-detected
  galaxies in the A3558 complex.
Note that while \citeauthor{Venturi00}'s sample is
  only very slightly deeper than ours in radio luminosity,
\citeauthor{Miller05}'s 
sample is much deeper and had its AGN removed (by rejection of optically
luminous galaxies).

For the A3528 complex, \cite{Venturi01} found a normal (respectively lack by a factor 2)
cumulative RLF of E/S0 galaxies in
the A3528 complex if the their fraction of early-type galaxies is 100\% (50\%).
But they saw no enhancement of the counts of
radio-detected galaxies in contrast with the strong excess found in the
optical.
Although our statistics are poor, there is a marginal excess of radio
luminous galaxies in the A3528 complex, as seen in the RLF of the top plot of
Figure~\ref{fig:ks}, and even in the wedge diagram of Figure~\ref{fig:wedges}.
Our results can be reconciled with the results of \cite{Venturi01}, only by assuming
that the shift is indeed not real as suggested by our marginal confidence level and that
nearly all the galaxies in the \citeauthor{Venturi01} sample are early-type,
which does not seem probable.

\cite{GJ86} found that optically-defined radio
  loudness was 3 times higher in clusters than outside clusters, but found no
  differences between their group, pair and isolated samples of radio-detected galaxies.
In comparison, we find that radio loudness is enhanced in clusters 
by typically 60\%
  outside of the SSC-CR, but decreased by a factor 3 inside the SSC-CR.
\cite{Best+07} find that AGN that are BCGs have the same distribution of
radio luminosities as non-BCGs. We find (Table~\ref{spearx200}) 
that radio luminosities of galaxies outside the SSC-CR
are little affected by the position in the cluster, while the SSC-CR galaxies
have radio luminosities that are anti-correlated with clustercentric
position, but that radio loudness is correlated with
clustercentric position: hence \emph{BCGs appear radio luminous, but given their
very high NIR luminosities, their radio loudness is low}.

There has been an accumulation of evidence over the last few years that
cluster-cluster mergers at the heart of the SSC are responsible for the
lack of radio-luminous galaxies in this region.
The discovery by \cite{Brand03} of a supercluster of radio galaxies is not in
contradiction with this result, if their supercluster has not yet reached a
stage of cluster-cluster merging, or if \citeauthor{Brand03} do not resolve
the supercluster core where cluster merging may occur.
Therefore, one has to be cautious 
when using radio-detected galaxies to trace large scale structures because the 
different dynamical stages of the collapse seem to play an important role in 
the presence of radio emission in galaxies.

\section{Physical processes at work in the core of the Shapley Supercluster}
\label{sec:process}

There are several physical processes at work in a cluster-cluster merger that
can affect the radio output of galaxies. 

The suppression of radio emission in cluster
galaxies within the SSC may be the result of cluster collisions, which heat up the
central regions of clusters and destroy their cool cores. Indeed,
\cite{FANM86} suggested that cooling flows fuel AGN and
\cite{Burns90} found that the probability that a cluster cD galaxy hosted an
AGN was 3 times greater (75\% vs. 25\%) if the cluster has a cool core.
Since A3558 has
no cool core \citep{SPO06}, one expects the central galaxy of A3558 to have
low radio loudness, as we observe.
On the other hand, in X-rays, A3528 is made of two components,
  both with cool cores \citep{Gastaldello+03}, which is consistent
with the high radio loudness of its cD.
We are not aware of measurements of
X-ray temperature profiles  of the other clusters with central
cDs: A3556, A3559, and A3566.

However, the reduced radio loudness in SSC-CR clusters is also present once
the central regions are removed from the analysis (last line of
Table~\ref{tab:KSinout}), even though its significance is reduced. 
So, \emph{while cool cluster cores might be responsible for suppressing the
  radio loudness of central cluster galaxies in the SSC-CR,
the suppression of radio emission also
  occurs outside the cluster centers}.

The AGN mechanism is believed to rely on three ingredients: the presence of a
massive black hole in the center of a galaxy, the supply of gas onto this
black hole, and a mechanism to transport this gas to the central black
hole. Starbursts also rely on the supply of gas onto molecular clouds.

Major mergers (of comparable mass galaxies) 
lead to violent relaxation that causes a substantial fraction of orbits to fall to 
the central regions and fuel the AGN \citep{Roos81}, as well as to compress the giant 
molecular clouds and induce a short but strong burst of star formation \citep{JW85}. 
Rapid non-merging galaxy collisions (hereafter flybys) also have an important effect 
on galaxies: the tidal field at closest approach will generate barred instabilities
\citep{GCA90}, which in turn will lead to efficient angular momentum exchange of stars 
and gas, some of which end up fueling the central AGN. Hence both galaxy mergers and 
rapid flyby collisions tend to boost the radio power of galaxies. Moreover, as we 
will discuss in Sect.~\ref{subsec:rps}, the ram pressure that galaxies feel when 
orbiting within the hot intracluster gas can be altered within colliding clusters.
One should note that these physical processes will affect the \emph{relative}
importance of the AGN/starburst activity relative to the mass of its host
galaxy, which are measured by our radio-loudness ${\cal R}_K$.

We hereafter provide a thorough argumentation of the relative importance of
several physical processes (major galaxy mergers, rapid galaxy flybys and ram
pressure stripping of interstellar gas in galaxies by the intracluster gas)
all believed to occur during a cluster-cluster merger. We discuss the impact
of these processes on the AGN and SBGs, taking the A3558 cluster complex as
an example.

\subsection{Increased or decreased direct major galaxy mergers within
  merging clusters?} 
\label{subsec:mergers}

Much insight can be found in the simple case of two equal mass
clusters merging in a head-on collision.  

Consider first the situation at
first pericenter, when the two clusters overlap.
The rate at which a test galaxy merges with other galaxies of comparable
mass is the sum of the major merger rate with galaxies from the 
first cluster and that with galaxies from the second cluster. 
Now, if the test galaxy originated in the first cluster,
the rate of mergers with galaxies from the first cluster will be the same as
when the two clusters were far apart. 

In the limit where the cluster-cluster
relative velocity is large, the test galaxy will suffer mergers with galaxies
from the second cluster at a negligible rate, since the collisions will be
too rapid to lead to galaxy mergers. In this regime, the test galaxy merges
with other galaxies at the same rate as before the cluster-cluster encounter.
For lower relative cluster-cluster velocities, the rate of mergers of the
test galaxy with galaxies of the second cluster will no longer be negligible,
hence the overall galaxy merger rate will be larger than it was before the
cluster-cluster collision. Hence, there will be more occasions 
where the galaxy gas will be funneled down to the inner regions and fuel the central
engine, thus triggering the AGN activity. And there will also be more
occasions to tidally compress the molecular clouds and generate starbursts.

In appendix~\ref{apprate}, we compute more precisely the rate of direct
major galaxy mergers with galaxies of the other cluster at the moment of the
first 
cluster overlap. For the SSC, the rms relative velocity of the clusters is
small in comparison with their internal velocity dispersions.
\begin{figure}[ht]
\centering
\includegraphics[width=\hsize]{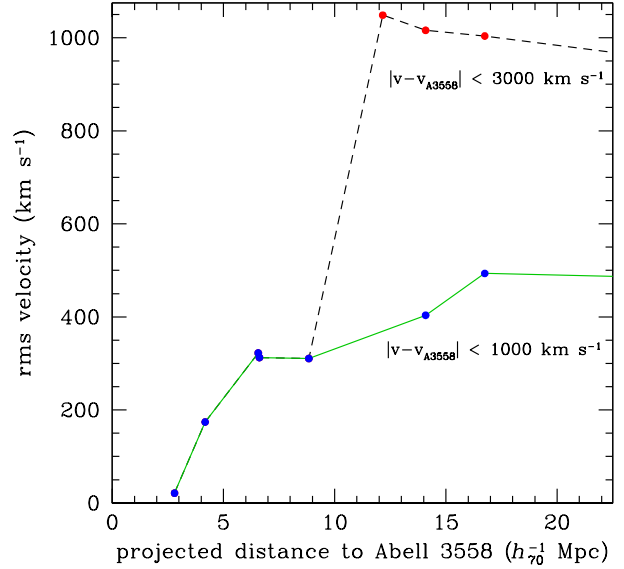}
\caption{Aperture velocity dispersion versus projected radius of the A3558 complex,
using two velocity cuts to remove foreground/background interlopers. 
Each \emph{filled circle} corresponds
  to the next closest cluster in projection to A3558, so the first symbol on
  the left is for 2 clusters, the next for 3 clusters, and so on.}
\label{sigap}
\end{figure}
Indeed, as can be seen in Figure~\ref{sigap},
the velocity dispersion is very low, $311 \, \rm km \, s^{-1}$ for
a projected apertures of $10 \, \rm Mpc$. However, 
the characteristic
velocity relative to A3558 will be $\sqrt{3}$ times larger, i.e. $538 \,
\rm km \, s^{-1}$. 
The rms internal velocity dispersion of the clusters within a projected
distance of 
$10 \, \rm Mpc$ of A3558 is $786 \, \rm km \, s^{-1}$.
Then, according to Figure~\ref{ratfig}, for luminous galaxies of velocity
dispersion $\sigma_g = 200 \, \rm km \, s^{-1}$ inside rich clusters of
velocity dispersion $\sigma = 800 \, \rm km \, s^{-1}$ (solid curve), with
$V/\sigma = 538/800=0.67$, we obtain an enhancement of the merger rate over 74\%.
Hence, the total merger rate is almost twice that in the isolated cluster.
For the merger rate of smaller galaxies, e.g. $\sigma_g = 100 \, \rm km \,
s^{-1}$ (dashed curve), the enhancement is still over 67\%.

But the moment of cluster overlap is very short, hence may not be so relevant.
After the first passage, the clusters will continue their course up to their
apocenter, at which point the merger rate of galaxies of the first cluster with
those of the second cluster will be near 0, as the two clusters are far
removed. The two clusters will then fall back onto one another for the final
merger (see \citealp{Barnes88} for the general case of merging galaxies).  

The maximum separation of clusters having crossed one another is
  comparable to the maximum distance a galaxy can reach after having crossed
  right through the center of a cluster. This latter value has been estimated
  to be between 1 and 2.5 times the 100 overdensity radius
  \citep{MSSS04,GKG05}, which converts to roughly 1.3 to 3.2 times $r_{200}$.
In fact, two equal mass clusters should feel a softer encounter than a tiny
  galaxy falling into a cluster, so that the maximum separation will be
  smaller (in relative terms). Now, the two clusters closest to A3558 --- A3556
  and A3562 --- lie at projected separations that correspond to 1.2 and 1.8
  times $r_{200}$. Of course the separations in real (3D) space are greater
  or equal to these projected separations. Nevertheless, it is possible that
  A3556 and/or A3562 have already crossed through A3558.

In roughly one cluster crossing time, violent relaxation will lead
to a relaxed merged cluster, and its galaxy major merger rate will scale as
$1/\sigma_v^3$ \citep{Mamon92}, which scales as $1/M$, given the cosmological
$M \propto \sigma_v^3$ relation applied to clusters. 
Hence, relative to the galaxy merger rate in
each of the progenitor clusters, the total rate of galaxy mergers will rise
from 1 at initial times to $\simeq 1.7$ at first pericenter, then back to
$\sim 1$ at apocenter, and will tend to 0.5 once the two clusters have merged and
violently relaxed (into a cluster twice their original mass).  
In other words, in an environment such as a supercluster,
with ongoing cluster mergers, the galaxy merger rate in clusters will be
boosted by a factor up to (during a short time at first pericenter)
$1.7/0.5=3.4$ relative to galaxy merger rate in an isolated cluster of the
same mass. 

In the case of an off-axis cluster-cluster merging encounter, the boost in
galaxy merger rates at first pericenter will be less than 3.4.
But, since the
galaxy merger rate 
in each of the two clusters will be at least greater than their original
merger rate, the boost relative to the final merged cluster will be greater
than $1/0.5=2$.

We now consider in more detail the clusters at some intermediate epoch
after the first pericentric passage and before the final merger.
It is well known that the tidal forces during an encounter convert
the orbital energy of the colliding pair into internal motions that deform
the systems.
How will this deformation affect the rate of galaxy mergers? Some of the
internal energy gained at the expense of orbital energy will be potential
energy.  
To first order the deformation is the sum of a monopolar deformation that
puffs up each cluster and higher order multipoles, for example possible tidal tails. 
These higher order terms have no incidence on the mean density. 
So the
net effect of the gain in potential energy (the loss of potential energy in
absolute value) will be 
a decrease in the number density of galaxies and a reduction of the
rate of galaxy collisions.
Moreover, part of the orbital energy  will be
converted into internal kinetic energy,
leading to an increased velocity dispersion of the galaxy system, hence a
lower fraction of collisions that will be slow enough to lead to mergers.
So, increases in both the potential and kinetic components of the internal
energy will combine to a decreased rate of galaxy mergers.

In other words, calling the potential and kinetic energies $W$ and $K$, using
primes for the puffed up state and writing $R'=R/x$ (where $x<1$), one has
$W'/W=x$, $\Delta K = -\Delta W = (1-x) W$ (from conservation of energy),
$K=-2 W$ (virial theorem), and $K'/K = (\sigma_v'/\sigma_v)^2$.
This leads to 
\begin{equation}
{\sigma_v' \over \sigma_v}=\sqrt{1+x\over 2} \ .
\label{sigratio}
\end{equation}  
As the
merger rate can be written as $n\,k$, where $k \propto \sigma_v^{-3}$ and 
$n$ is the galaxy 
number density
(\citealp{Mamon92}, and eq.~[\ref{dNdt}]), the ratio 
of the galaxy merger rates between the puffed up cluster and the normal one
is found to be (using eq.~[\ref{sigratio}]) 
 \begin{equation}
{n'\,k' \over n\,k} = 2^{3/2} {x^3 \over (1+x)^{3/2}} \ .
\label{mergrate}
\end{equation}

When one
considers systems including a dissipative gas component, the global effect of
the energy transfer is small on the non-dissipative component. While we are
interested in the galaxy component of clusters, such simulations have not yet
been performed in a realistic manner. 
However, the dark matter component, which should have very
similar dynamics to the galaxy component, seems very little affected by the
encounter. Indeed, as can be seen in the snapshots of merging clusters of
\cite{RBL93} and of merging galaxies of \cite{dMCMS07}, the puffing up of the
non-dissipative component appears to be of order of or less than 20\%, i.e. a
factor $x \geq 0.8$.
So to explain the factor two decrease in the fraction of radio-loud
galaxies 
in the A3558 cluster complex (Sect.~\ref{sec:radiofracs}), using
equation~(\ref{mergrate}) one would
therefore require $x = 0.74$, i.e. a 35\% increase in the cluster sizes,
which appears to be inconsistent with the simulations quoted above.

To summarize, \emph{at first passage of the merging clusters, the rate of
  direct galaxy mergers is increased, while  subsequently, before the final merger
  of the two clusters, the rate of galaxy mergers is decreased.}
Since the factor two reduction in the
fraction of radio-loud galaxies in A3558 cluster
  complex requires an unrealistic 35\% puff up in the clusters, we conclude
  that \emph{galaxy mergers cannot explain by themselves the reduced loudness in
  cluster radio-detected galaxies inside the SSC-CR.}

\subsection{The effects of rapid flyby collisions}
\label{subsec:flybys}

Naturally, rapid flybys are less efficient individually than mergers in
inducing star formation and AGN activity.  But, in rich clusters, flybys are
much more common than galaxy mergers.  \cite{Mamon00_IAP} computed
analytically both the rate of major mergers and the rate of flybys as a
function of galaxy environment, mass and position in its group or cluster
environment.  His Figure 5 indicates a rate of flybys (defined such that they
pump in an amount of energy at least one-third of the binding energy of the
test galaxy) that, for rich clusters, is about 1 per galaxy per Hubble time,
independent of galaxy mass.  In comparison, ongoing galaxy merger rates in
present-day rich clusters are almost always at rates less than 0.1 per
galaxy, extrapolated to a Hubble time. So, \emph{in rich clusters, strong
flybys are about 10 times more common than major mergers}.

The rate of flybys is enhanced during the overlapping part of the cluster
encounter. According to equation (19) of \cite{Mamon00_IAP}, the rate $k$ of
flybys varies as $1/\sigma_v$, hence as $1/M^{1/3}$.  Just like for galaxy
mergers, the total rate $n\,k$ of flybys of galaxies from either cluster will be
$\la$ twice that in the individual progenitor clusters
(without embarking into a calculation similar to that of
appendix~\ref{apprate}), then will fall to below unity 
(see Sect.~\ref{subsec:mergers})
once the two clusters
separate towards their apocenter, increases again at the final cluster
merger, but then decreases after the final relaxation to reach a value of
$2^{-1/3}$ the flyby rate in the individual clusters.  The boost in the flyby
rate relative to that in relaxed isolated clusters of the same mass will thus
be $\la 2^{4/3} \simeq 2.5$ at first pericenter.
For off-axis cluster-cluster encounter, the flyby rate at first pericenter
will still be greater than that in the individual progenitor clusters, hence
the boost of the rate of flybys at pericenter, relative to that in isolated
clusters of the same mass will be $2^{1/3} \simeq 1.3$.

Once the clusters pass their pericenter and puff up through their mutual
tides (see Sect.~\ref{subsec:mergers}), the rate of flyby encounters will be
decreased.
With the $1/\sigma_v$ scaling of the rate of flybys, and 
using equation~(\ref{sigratio}), we find that the ratio of the rate of flybys
between the puffed clusters past pericenter and the individual clusters will
be
 \begin{equation}
{n'\,k' \over n\,k} = 2^{1/2} {x^3 \over \sqrt{1+x}} \ .
\label{flybyrate}
\end{equation}
Equation~(\ref{flybyrate}) then implies that to explain the factor of two
decrease in the fraction of radio-loud galaxies in the A3558 cluster complex,
one requires $x=0.78$, i.e. a 28\% increase in the cluster sizes,
which again  
appears to be inconsistent with the hydrodynamic simulations of \cite{RBL93}.

In summary, flybys are enhanced during the pericentric passage, and decreased
afterwards, when the clusters puff up by their mutual tidal interaction. Yet,
\emph{flybys cannot explain by themselves the factor two decrease in the
  fraction of radio loud galaxies in the core of the SSC.}

\subsection{Increased ram pressure stripping of galaxies in merging clusters} 
\label{subsec:rps}

We now argue that in merging clusters, as is the case in the inner regions of
the SSC \citep{Bardelli98},
ram pressure stripping of interstellar
gas will be considerably more effective than in single clusters, which
should
lead to the starvation of the AGN activity as well as of star formation, and 
might thus explain the low levels
of radio loudness for  radio-detected galaxies within a virial radius of 
SSC-CR clusters, relative to those within a  virial radius of clusters outside
of the SSC-CR (Fig.~\ref{fig:corrsel} and Table~\ref{tab:KSinout}).

When two clusters merge, their gaseous intracluster media, in contrast to
their stellar and dark matter components, will not interpenetrate, but
instead dissipate their orbital energies in a strong shock, caused by the adiabatic
compression of the colliding gas clouds \citep*{RBL93}.
Therefore, a shock-heated interface should lie between the two clusters,
reminiscent of what is seen in Stephan's Quintet (the HCG 79 compact group of
galaxies, \citealp{vdHR81,PTAS97}) and seen in all realistic hydrodynamical
simulations of merging clusters (e.g. \citealp{RBL93}).
Behind the shock, the gas component should follow the same motion as the mean
motion of the galaxies. The galaxies of the first cluster will see the
relative speed of the gas change as they move from their initial parent
cluster, through the shock to the other cluster.
Since ram pressure is $P=\rho_{\rm gas}\,v^2$, 
where $\rho_{\rm gas}$ is the gas density and $v$ the velocity of
the galaxy relative to the intracluster gas  \citep{GG72}, the
efficiency of the ram pressure stripping will vary according to whether a
galaxy is in it's initial parent cluster, in the shock at the cluster-cluster
interface, or in the second cluster.

Consider two clusters of equal mass, with 1D velocity dispersion $\sigma$,
merging along the $x$ axis at relative velocity $V$.  At the moment it passes
through the shock, the shocked gas that the galaxy encounters is 4 times
denser than before the shock. Moreover, if the galaxy travels along the $x$
axis, the wind that the galaxy feels will typically be enhanced by
$1+(V/2)^2/\sigma^2$.  Hence a galaxy traveling along the $x$ axis, passing
through the shock interface, will feel a ram pressure $P'$ enhanced by
\begin{equation}
{P'\over P} = 4 + \left ({V\over \sigma}\right)^2 \ .
\label{enhanceshock}
\end{equation}
Galaxies traveling in perpendicular directions will feel an
enhanced ram pressure of a factor 4, and their internal gas will be
efficiently stripped by the ram pressure of the intracluster gas at the
shock.

After the galaxy, traveling along the $x$ axis, passes through the shock,
it will ``see'' the second cluster moving towards it at a velocity equal to
its velocity relative to the first cluster plus the cluster-cluster relative
velocity.
Hence, its interstellar gas will be ram pressure stripped
by the intracluster gas of the other cluster much more efficiently, by a
factor 
\begin{equation}
{P'\over P} = 1+\left ({V\over\sigma}\right)^2 \ .
\label{enhancepostshock}
\end{equation}
Note that a galaxy moving within the first cluster along the $x$
axis, but in the same direction as the second cluster, will feel less ram
pressure stripping during its initial infall and subsequent rebound.
However, during its recoil\footnote{The recoil timescale, $\approx
      R_{\rm gal}/\sigma_v$ 
should be shorter than the cluster-cluster merger time, $\approx R_{\rm
  cluster}/V$, since the cluster radius $R_{\rm cluster}$ is by definition
larger than the radial position of the galaxy $R_{\rm gal}$ and $V < 538 \,
\rm km \, s^{-1} < \sigma_v \approx 800 \, \rm km \, s^{-1}$.}
 it will find itself in a similar situation as the
galaxy that initially moves in the direction of the second cluster, and at
this stage it will experience strong ram pressure stripping.

Galaxies traveling along axes perpendicular to the
$x$ axis will feel the same ram pressure as if there had been no
  cluster-cluster merger.
Therefore, within a merging cluster system, some of the galaxies that have penetrated
the second cluster will suffer
additional ram pressure stripping while others will feel about the same ram
pressure as in their initial cluster.

Equations~(\ref{enhanceshock}) and (\ref{enhancepostshock})  suggest
that for galaxies traveling along the cluster-cluster collision axis, 
the effect of the shock is more dramatic than the
effect of the enhanced wind of the second cluster.
Given the typical velocity dispersion of $800 \, \rm km \, s^{-1}$ for clusters in the A3558
complex, with $V=538 \, \rm km \, s^{-1}$, one finds that a galaxy moving
along the cluster-cluster collision axis will see a ram pressure enhanced by
a factor of 4.45 at the shock (eq.~[\ref{enhanceshock}]) and by a factor 1.45
past the shock (eq.~[\ref{enhancepostshock}]). 
However, while the mild enhancement of ram pressure past the shock is long-lived,
the thickness of the shock front is small in
  comparison with the cluster size \citep{RBL93}, so that the galaxy will
  feel an enhanced density only for the short time it crosses through the
  shock. Therefore, it is not clear whether the short time during which the galaxy
  feels a ram pressure enhanced by a factor 4 will be sufficient for the
  interstellar gas to escape its host galaxy. 
It is beyond the scope of the present paper to predict which of the shock or enhanced
wind from the 2nd cluster is the most efficient at decreasing the gas content
of galaxies in merging clusters.

Note that the very low velocity difference, $21 \, \rm km \, s^{-1}$, between
A3558 and its nearest projected neighbor A3556 suggests that these two
clusters have already passed through each other (unless their relative
motion is perpendicular to the line of sight) and lost a substantial
fraction of their initial orbital energy. The extra wind from the second
cluster (A3556) will presently 
be of the same order as that from the first cluster (A3558). But before
the two clusters lost their mutual orbital energy, the second wind must have
been roughly twice as fast as the first one.
 
Moreover, ram pressure will strip the low density atomic gas clouds more
  efficiently than the dense cores of giant molecular clouds. The latter will
  be compressed by the ram pressure \citep{Bekki03}. 
Therefore, if the galaxies feel a burst
  of ram pressure as they travel through the shock front, they should experience
  a burst of star formation and hence an enhancement of their radio
  emission. But once the atomic gas is removed, the 
  subsequent star formation is starved. These issues are complex, 
  probably require high resolution hydrodynamical simulations, and are
  beyond the scope of the present paper.

\subsection{Synthesis}
\label{subsec:discussion}

We have argued that, on one hand, environments such as the SSC which seem to
harbor ongoing cluster-cluster merging collisions, should enhance the rates of
direct galaxy-galaxy mergers (Sect.~\ref{subsec:mergers}) and rapid
flyby collisions  
(Sect.~\ref{subsec:flybys}) at the first passage of the
two clusters. But both rates should be decreased after the 
first pericentric passage and before the final cluster-cluster coalescence. 
However, neither galaxy mergers nor rapid flybys appear sufficient to explain
the factor of two decrease in the fraction of radio loud galaxies in the A3558
complex at the core of the SSC.

On the other hand, we argued in Sect.~\ref{subsec:rps}
that both the enhanced density at the
shock front of the cluster-cluster collision and the enhanced wind felt when
galaxies pass through this shock ought to
enhance the efficiency of
the stripping of interstellar gas by the hot intracluster gas, which in turn
will starve AGN and star formation.

Also, the enhanced ram pressure stripping 
should affect more than half the cluster galaxies that will
have crossed the shock front (the orbital timescale within the cluster will
be considerably shorter than the cluster-cluster merging time). 
On the other hand,
only a few percent of cluster galaxies are expected to merge
for the few Gyr duration of the cluster-cluster merger.
Indeed, Figure~4 of \cite{Mamon00_IAP}, shows that, at best, the current
rate of galaxy mergers in clusters is of order of 0.14 per Hubble time,
i.e. of order of 0.01 per Gyr.
Hence, a reduction of a factor 2 to 3
in the galaxy merger rate after the first cluster-cluster
pericentric passage will decrease the fraction of galaxies that merge
during the last 3 Gyr from roughly 3\% to 1\%.
If the bursts of star formation in clusters are mostly caused by 
infalling spirals from low velocity dispersion groups where the merger
rate is high, the effect of the cluster-cluster collision on the rate of star
formation should be negligible.

Moreover, while a decreased rate of galaxy mergers or flybys will explain the
decreased fraction of radio-loud galaxies, \emph{only enhanced ram pressure 
stripping in merging clusters can explain that the few radio-detected
non-BCG galaxies within SSC-CR clusters have values of radio loudness typically
several times lower than  
radio-detected galaxies within clusters outside of the SSC-CR}.
Therefore, in an environment of merging clusters, it is likely that galaxies
will have decreased radio loudness from the starvation of AGN and star
formation activity caused by the enhanced ram pressure stripping of galaxies,
unless the galaxies are in clusters that are just approaching for the first
time, as appears to be the case for the A3528 cluster complex.
\emph{The anti-correlation of radio loudness with the density of the 
environment that we find in the core of the Shapley supercluster suggests 
that clusters in this environment are going through a pericentric
  passage now, or have done so
in the recent past}.

The evolution of radio-detected galaxies should therefore be
linked to the dynamical  
evolution of structures and to the merging of the different sub-structures, 
thus potentially explaining the apparent inconsistencies between 
various works (e.g. \citealt{Owen99} and \citealt{Venturi00}). In addition, with less 
massive (and lower velocity dispersion) structures in the past, 
there would be more galaxy merging 
and less ram pressure stripping in cluster-cluster merging that
would not quench star formation and the AGN phenomenon as well as it does
nowadays. The radio-detected galaxies could therefore very well remain good tracers
of large scale structures at high redshift (as found by \citealt{Brand03}).

\section{Summary}
\label{sec:sum}  

We merged two of the most recent optical/NIR catalogs (FLASH and 6dFGS) that 
cover a large superstructure (the Shapley Supercluster) as well 
as a fair number of clusters and voids. 
We then cross-identified our sample with the NVSS radio survey and restricted our 
study to a volume, flux and radio luminosity limited sample to limit
biases in our studies. We retrieved 142
radio-detected galaxies, from which 28 were classified as AGN and 114 as SBGs.

To first order, radio-detected galaxies roughly trace the overall large scale 
structure of the area (Figs.~\ref{fig:wedges} and
  \ref{fig:numgals}). However,  
the fraction of galaxies with high $K$-band normalized
radio-loudness in the A3558 cluster complex at the core of the SSC is half
that in our reference sample (see Fig.~\ref{fig:ks}).
Moreover, we find that radio-loudness is clearly anti-correlated with the
large scale density of the environment (Fig.~\ref{fig:cutoutL}).
This anti-correlation disappears when a large shell around the central region of 
the supercluster (SSC) is removed, indicating that this region is mostly 
responsible for this trend. 

A detailed analysis indicates that this radio loudness / density
anti-correlation is caused by the lower radio loudness of in-cluster
($r\!<\!0.8\,r_{200}$) radio-detected galaxies within the
$10\,\rm Mpc$ region  
centered on A3558 (the center of the SSC), in comparison with the
in-cluster radio-detected galaxies outside the core of the SSC
(see Fig.~\ref{fig:corrsel}).
The suppression of radio loudness in radio-detected galaxies appears most
dramatic for BCGs in the SSC-CR (Fig.~\ref{fig:loudvspos}). But the
suppression of radio loudness is also significant in non-BCG cluster radio
galaxies, especially at $R < 0.3\,r_{200}$.
While NIR luminosity segregation is extremely strong in the SSC-CR, it is
also significant outside the SSC-CR (Table~\ref{spearx200}), 
so that the modulation by the SSC-CR of
the distribution of
radio loudness of in-cluster galaxies is at least as significant as 
the corresponding distribution
of NIR luminosities (Table~\ref{tab:KSinout}).

The radio luminosity distributions in the A3558 
and A3528 cluster complexes 
show
(Fig.~\ref{fig:ks}) different behaviors that might highlight different dynamical states. 
The radio-detected galaxies in the A3558 complex have no highly  radio-luminous galaxies 
and are significantly less radio-loud than in our reference sample, 
while those in the A3528 complex are significantly more radio-luminous 
and marginally more radio-loud.

In light of these results, we investigated 
possible causes for the suppression of radio loudness in cluster
  galaxies of the SSC core region, in the context of cluster-cluster 
mergers, which are thought to occur in this region.
For the BCGs, a likely culprit is the disruption of the cool cluster core,
which is believed to provide an efficient mechanism to fuel the AGN.
For non-BCGs, using simple calculations, 
we show (Sect.~\ref{sec:process}) that both direct major galaxy mergers
and rapid collisions are 
enhanced at the moment of cluster overlap
in comparison with isolated clusters of the same mass.
However, after their first passage, the two colliding clusters will puff up
(through their mutual tidal interaction), which leads to decreased rates of
direct galaxy mergers and flybys. But we show that the cluster puff up is too small
to explain the decreased fraction of radio-loud galaxies in the core of the
SSC through galaxy mergers and rapid flybys.
On the other hand, ram pressure is increased 
 by galaxies passing
through the shock front at the cluster-cluster interface, and also by the
enhanced wind velocity they feel as they enter the second cluster.
This enhancement of ram pressure will be most effective in the inner
regions of clusters and 
should quench the AGN activity and starve the star formation, thus reducing
the radio loudness of cluster galaxies in the SSC core, as is observed.

There are many perspectives to this work.
The range of radio luminosities or radio loudness values where the
suppression of radio emission occurs can be better assessed by 
studying a deeper radio sample of the SSC-CR, such as that of
\cite{Miller05}.
The physical mechanisms at work can also be better constrained by
distinguishing AGN from SBGs,
using optical spectra 
or far IR diagnostics, and by
incorporating galaxy morphologies to the analysis.

For example, one could test the effects of cluster-cluster
  mergers using the colors or 
  spectra of galaxies: these  
will show a burst of star formation, whose age reflects the time of the first
pericentric passage, since starburst timescales are much shorter than the 
dynamical timescales of structure merging. 
The variation of the star formation rate with time
around pericenter should be skewed towards early times. Indeed, the passage of
a cluster galaxy through the shock front and into the faster wind caused by
the motion of the second cluster will abruptly starve the star formation that
had just been enhanced by increased galaxy mergers at the beginning of the
penetration of the two clusters. 

Also, it would be interesting to see if the SSC-CR is unique or common, by
performing similar analyses as done here on the core regions of other nearby superclusters.
Finally, the details of ram pressure stripping, in particular the competition
between the short but strong enhancement caused by passage of galaxies
through the shock front generated by the cluster-cluster merger on one hand, and
the faster wind felt after passage through the shock on the other hand, can be
assessed by 
analyzing currently available hydrodynamical $N$-body simulations of cluster
mergers.

\vskip 1cm
 
\begin{acknowledgements}
We thank 
Andrea Biviano,
Avishai Dekel, 
Niruj Mohan,
Nick Seymour, 
and especially Trevor Ponman 
for enlightening discussions. 
We are indebted to Gwena\"el Bou\'e for help with
the Appendix, as well as to Andrea Biviano, Florence Durret,
Gary Hill, Steve Rawlings for a 
critical reading of temporary versions of the manuscript. 
J.C.M. is grateful to Gary Hill for suggesting the topic and for 
hosting him at the University of Texas during the early stages of 
this work, which were supported, in part, by the Texas Advanced
Research Program under Grant No. 009658-0710-1999.
We also thank an anonymous referee for his numerous comments, which led
to a significantly improved work.
This publication makes use of data products from the Two Micron All Sky
Survey, which is a joint project of the University of Massachusetts and the
Infrared Processing and Analysis Center/California Institute of Technology,
funded by the National Aeronautics and Space Administration and the
National Science Foundation.
The Digitized Sky Surveys were produced at the Space Telescope Science
Institute under U.S. Government grant NAG W-2166. The images of these
surveys are based on photographic data obtained using the Oschin Schmidt
Telescope on Palomar Mountain and the UK Schmidt Telescope.
This research has made use of the NASA/IPAC Extragalactic Database (NED)
which is operated by the Jet Propulsion Laboratory, California Institute of
Technology, under contract with the National Aeronautics and Space
Administration.

\end{acknowledgements}

\appendix

\onecolumn

\section{Relation between virial radius and aperture velocity
  dispersion for an NFW model} 
\label{app:dynamical}  

Given that the mass $M_{\rm vir}$ within the virial radius $r_{\rm vir}$ is
$M_{\rm vir} = {4\pi/3} \Delta \rho_c \,r_{\rm vir}^3$,
where $\Delta$ is the mean density of the cluster within the virial radius,
relative to the critical density $\rho_c = 3 H^2(z)/(8\pi G)$, 
one easily derives
\begin{equation}
\sigma_{\rm ap} = \left ({\Delta\over2}\right)^{1/2}\,\left
({\sigma_v\over v_{\rm vir}}\right) 
\,H(z)\,r_{\rm vir} \ ,
\label{sigvth}
\end{equation}
where $v_{\rm vir} = (G M_{\rm vir}/r_{\rm vir})^{1/2}$ is the circular velocity at the
virial radius.

Inspecting Figure~7 of \cite{LM01}, one finds for an isotropic 
\citeauthor*{NFW96} (\citeyear{NFW96}, hereafter NFW) model that the
\emph{normalized aperture velocity dispersion} 
$\widetilde \sigma_v = \sigma_{\rm ap} / v_{\rm vir} \simeq 0.7$. 
One can precisely compute $\sigma_{\rm ap}$ for isotropic velocities
with a simplified version of
equation~(47) of \cite{LM01} given by \cite{ML05a}:
\begin{eqnarray}
\sigma_{\rm ap}^2 (R) &=& {4\pi\,G\over \textcolor{red}{3}\,M_p(R)}\,\left [\int_0^\infty r\,
\rho(r) M(r)\,{\rm d}r - \int_R^\infty  {(r^2\!-\!R^2)^{3/2}\over r^2}\,\rho(r)
M(r)\,{\rm d}r \right ] 
\ ,
\label{sigapisogen}
\end{eqnarray}
where $\rho$, $M$, and $M_p$ are the tracer mass density, total and
projected mass, respectively (the \citeauthor{LM01} formula involved a
double integral instead of the two single ones in eq.~[\ref{sigapisogen}]).
Writing the density and mass of the NFW profile of virial radius $r_{\rm vir}$ and
concentration $c$ as
\citep{ML05a}
\begin{eqnarray} 
\rho(r) &=& {1\over g}\,{M_{\rm vir}\over 4\pi r_{\rm vir}^3}\,\widetilde \rho \ ,
\label{rhoofr}
\\
M(r) &=& M_{\rm vir}\,{\widetilde M\over g} \ ,
\label{Mofr}
\end{eqnarray}
where
\begin{eqnarray}
\widetilde \rho &=& y^{-1}\,(1+y)^{-2} \ ,
\label{rhoofy}
\\
\widetilde M(y) &=& \ln(y+1) - {y\over y+1} \ , 
\label{Mofy}
\\
y &=& {r\over a} = {c\,r\over r_{\rm vir}} \ ,
\label{ydef}
\\
g &=& \widetilde M(c) = \ln(c+1) - {c\over c+1} \ ,
\label{gdef}
\end{eqnarray}
(where our definition of $g$ is the inverse of that of \citeauthor{LM01} and
where $a$ is the radius of density slope $-2$),
equation~(\ref{sigapisogen}) yields 
\begin{equation}
\widetilde \sigma_v^2 (Y\,r_{\rm vir}/c) = {c\over \textcolor{red}{3}\,  g}\,{\int_0^\infty y \,\widetilde \rho
\,  \widetilde M\,dy - \int_Y^\infty \left (y^2-Y^2\right )^{3/2}\,
\widetilde \rho \,
  \widetilde M\,dy/y^2
\over
{\cal C}^{-1}(1/Y) / \left |Y^2-1\right|^{1/2} + \ln(Y/2)} \ ,
\label{sigaptilde}
\end{equation}
where $Y = R/a = c\,R/r_{\rm vir}$ and
\[
{\cal C}^{-1}(x) = \left \{\begin{array}{ll}
\displaystyle
\cos^{-1} x& \qquad (x<1) \ , \\
\displaystyle
\cosh^{-1} x& \qquad (x > 1) \ ,
\end{array}
\right.
\]
(see eq\textcolor{red}{s}.~[42] \textcolor{red}{and [43]} of \citealp{LM01},
note that the expression in brackets in 
eq.~[43] of \citealp{LM01} tends to $1-\ln 2$ for $Y=1$).
Inserting equations~(\ref{rhoofy}), (\ref{Mofy}) and (\ref{gdef}) into
equation~(\ref{sigaptilde}), yields a normalized aperture velocity dispersion
at the virial radius ($Y=c$)
\begin{equation}
{\sigma_{\rm ap}\over v_{\rm vir}} = 
\widetilde \sigma_v \simeq \hbox{dex} \left (-0.1539 - 0.2138\,\log c +
0.2358\,\log^2 c - 0.05357\,\log^3 c  + 0.005515\,\log^4 c \right ) \ ,
\label{sigapovervvvsc}
\end{equation}
which is accurate to better than 5\% for $c>0.5$ and better than 0.12\% for
$c>1$. 
Thus, we find that 
$\widetilde \sigma_v$ reaches a minimum of 0.62 between $2 < c
< 5$ and is equal to 0.66 for $c=10$.

With 
$H_0 = 70 \,\rm km \,s^{-1} \, Mpc^{-1}$
and $\Delta=200$, equations~(\ref{sigvth}) and (\ref{sigapovervvvsc}) lead to
\begin{eqnarray} 
\left ({r_{200}\over
  1\, \hbox{Mpc}}\right ) = 
{\sigma_{\rm ap} \over \sigma_1} \ ,
\qquad
\left ({\sigma_1\over 1 \, \rm km \, s^{-1}} \right ) = 
\left \{
\begin{array}{ll}
433 & (c=4)\\
436 & (c=5)\\
463 & (c=10) \\
\end{array}
\right.
\ .
\label{rvirsigv}
\end{eqnarray}

Note that \cite*{CYE97} assume $\sigma_{\rm ap}/v_{\rm vir} =
1/\sqrt{3}$, hence 
$\sigma_{\rm ap = \sqrt{\Delta/6}\,H(z)\,r_{\rm vir}}$,
which for $H_0 = 70 \,\rm km \,s^{-1} \, Mpc^{-1}$ and $\Delta = 200$ 
yields $\sigma_1 = 404 \,
\rm km \, s^{-1}$. In other words, relative to a $c=5$ NFW cluster,
\citeauthor{CYE97} overestimate the virial radius by $436/404-1 = 8\%$.

\section{Rate of direct galaxy mergers in fully overlapping 
merging clusters of equal mass}
\label{apprate}
The rate at which a galaxy suffers direct major mergers can be written
\begin{equation}
{dN\over dt} = n k_1 \ ,
\label{dNdt}
\end{equation}
where $n$ is the number density of galaxies, while the merger rate $k_1$ (`1'
for single cluster) is \citep{Mamon92}
\begin{equation}
k_1 = \left \langle v S(v) \right \rangle = \pi\,\int_0^\infty v\,f(v)\,p_{\rm
  crit}^2(v)\,dv \ ,
\label{k11}
\end{equation}
with $S(v) = \pi\,p_{\rm crit}^2(v)$ the merger cross-section and $f(v)$ the
distribution of relative galaxy velocities, normalized such that
$\int_0^\infty f(v) dv = 1$.
One typically assumes a Gaussian distribution of relative velocities, with
standard 
deviation $2^{1/2}\,\sigma$, where $\sigma$ is the one-dimensional cluster
velocity dispersion and the factor $2^{1/2}$ arises from the consideration
of relative velocities, yielding
\begin{equation}
f(v) = {1\over 2 \,\sqrt{\pi}\,\sigma^3}\,v^2 \exp\left (-{v^2\over 4
  \sigma^2}\right ) \ ,
\label{fofv}
\end{equation}
which satisfies $\int_0^\infty f(v)\,dv = 1$.

Equations~(\ref{k11}) and (\ref{fofv}) lead to
\begin{equation}
k_1 = {1\over 2\,\sqrt{\pi}\,\sigma^3}
\int_0^\infty v^3 \exp \left (-{v^2\over 4\,\sigma^2}\right)\,S(v)\,dv \ .
\label{k12}
\end{equation}
The direct merger rate $k$ was computed by \cite{Mamon92,Mamon00_IAP} for the
case of a linear critical impact parameter \citep{RN79}, yielding 
\begin{equation}
S(v) = \pi\,p_0^2\,\left (1-{v\over v_0} \right )^2 \ ,
\label{Sofv}
\end{equation}
where $p_0$ is the maximum impact parameter that leads to a merger, while
$v_0$ is the maximum velocity for head-on mergers.

We now consider the rate of direct galaxy mergers within a system of two
equal mass merging
clusters, at the moment when the two clusters overlap.
We call $V$ the relative velocity of the two clusters at overlap.
A galaxy initially in cluster 1, will suffer mergers with other galaxies of
cluster 1 at a rate $k_1$ given by equation~(\ref{k12}) and galaxies of
cluster 2 at a rate $k_2$, with a total merger rate
\[
{dN\over dt} = n_1 k_1 + n_2 k_2 = n\,\left(k_1 + k_2\right)\ .
\]
The rate $k_2$ of direct major mergers that the test galaxy will suffer with
galaxies of cluster 2 can be written as follows.

Consider a cylindrical coordinate system, whose axis joins the 2 clusters and
call $v_\parallel$ and $v_\perp$ the coordinates of a galaxy in this
coordinate system in the frame
of the 2nd cluster.
In analogy with equation~(\ref{fofv}), one can then express the
distribution of the velocities of the galaxies in the 2nd cluster as
\begin{equation}
f_2\left(v_\parallel,v_\perp\right) = {1\over 4 \sqrt{\pi} }\,{v_\perp\over
\sigma^3}\,\exp -\left ({v_\parallel^2+v_\perp^2 \over 4
    \,\sigma^2} \right ) \ ,
\label{f2}
\end{equation}
which verifies $\int_{-\infty}^{+\infty}dv_\parallel\!\int_0^\infty
f_2\left(v_\parallel,v_\perp\right) dv_\perp = 1$, and where $\sigma$ is the
velocity dispersion of the 2nd cluster (equal to that of the first cluster).
Noting that the velocity of a galaxy of the 2nd cluster relative to the test
galaxy of the first cluster satisfies
$v^2 = \left (v_\parallel+V\right)^2 + v_\perp^2$,
the analog of equation~(\ref{k11}) leads to
\begin{eqnarray}
k_2 &=& \int_{-\infty}^{+\infty} dv_\parallel \int_0^\infty
f_2\left(v_\parallel,v_\perp\right)\,v\,S(v)\,dv_\perp
\nonumber \\
&=& {\exp \left (V^2/\sigma^2\right) \over 4 \sqrt{\pi}\,\sigma^3}\,
\int_{-\infty}^{+\infty} \exp \left ({v_\parallel V \over 2\,\sigma^2} \right
)\,dv_\parallel
\,\int_{|V+v_\parallel|}^\infty v^2\,S(v)\,\exp \left (-{v^2
\over 4 \sigma^2} \right )\,dv \nonumber \\
&=& {1 \over 4 \sqrt{\pi}\,\sigma^3}\,
\exp \left ({V^2\over \sigma^2}\right)
\,
\int_0^\infty v^2\,S(v)\, \exp \left (-{v^2
\over 4 \sigma^2} \right )\,dv
\,\int_{-V-v}^{-V+v} \exp \left ({v_\parallel V \over 2\,\sigma^2} \right
)\,dv_\parallel \nonumber \\
&=& {1\over \sqrt{\pi}\,V\,\sigma}\,\exp\left (-{V^2\over 4\,\sigma^2}
\right)\,
\int_0^\infty v^2
\sinh \left ({v\, V\over 2\, \sigma^2}\right )
\, \exp \left (-{v^2\over 4\,\sigma^2}\right)\,S(v)\,dv \ ,
\label{k2}
\end{eqnarray}
where the 2nd
equality was found using the expression of $f$ of equation~(\ref{f2}),
and the 3rd equality by 
changing variables, writing $v_\perp\,dv_\perp\,dv_\parallel
= v\, dv \,dv_\parallel$.

For a linear decrease of $p_{\rm crit}$ with $v$ \citep{RN79}, 
i.e. $S(v) \propto
(v_0-v)^2$ (eq.~[\ref{Sofv}]), the ratio of the rate 
of mergers with galaxies of cluster
2 to that with galaxies of the test galaxy's own cluster is found by
integrating equation~(\ref{k2}) and dividing by the integral \citep{Mamon92}
of equation~(\ref{k12}):
\begin{eqnarray}
{k_2\over k_1} &=& 
{2\,\sigma^2\over V}\,\exp\left (-{V^2\over 4\,\sigma^2} \right )
\,
{\displaystyle
\int_0^\infty v^2
\sinh \left ({v \,V\over 2\, \sigma^2}\right )
 \exp \left (-{v^2\over 4\,\sigma^2}\right)\,S(v)\,dv
\over
\displaystyle
\int_0^\infty v^3
 \exp \left (-{v^2\over 4\,\sigma^2}\right)\,S(v)\,dv
}
\nonumber \\
&=& {\exp\left (-Y^2\right)\over 2 Y}\,
{\int_0^x u^2\,\sinh(2\,u\,Y)\,\exp(-u^2)\,(x-u)^2\,du
\over
\int_0^x u^3\,\exp(-u^2)\,(x-u)^2\,du
} \nonumber \\
&=& 
{\exp \left [-Y^2-x(x+2Y) \right ]\,
\over 32\,Y\,\left [1-\exp \left (-x^2\right)+x^2/2-(3/4)\sqrt{\pi} x \,{\rm
 erf}(x) \right ]} \,F(x,Y) \ ,
\label{rat}
\end{eqnarray}
where $u=v/(2\,\sigma)$,  $x=v_0/(2\,\sigma)$, $Y=V/(2\,\sigma)$, and
\begin{eqnarray}
F(x,Y)
&=&2\,\left \{
4\,\exp \left [x\,\left( x + 2\,Y \right) \right]\,x^2\,Y + 
     \left[\exp \left (4\,x\,Y\right) -1 \right] \,x\,
\left( 1 + 2\,Y^2 \right)   \right. \nonumber \\
&\mbox{}& -\,    
\left[ 1 + \exp \left (4\,x\,Y\right) 
- 2\,\exp \left [x\,\left( x + 2\,Y \right) \right] \,\right] \,
Y\,\left.\left( 5 + 2\,Y^2 \right)  \right\}  \nonumber \\
&+& 
{\sqrt{\pi }}\,  \exp\left [\left( x + Y \right)^2\right] \nonumber \\
&\times&
   \left\{ \left[ 3 + 4\,Y^2\,\left( 3 + Y^2 \right)  - 
        4\,x\,Y\,\left( 3 + 2\,Y^2 \right)  + 
        x^2\,\left( 2 + 4\,Y^2 \right)  \right] \,{\rm erf}(x - Y) \right.
\nonumber   \\
&\mbox{}&
     + \,2\,\left[ 3 + 2\,x^2 + 4\,\left( 3 + x^2 \right) \,Y^2 + 4\,Y^4 \right]
        \,{\rm erf}(Y) \nonumber \\
&\mbox{}& \times \left.  
     \left[ 3 + 2\,x^2 + 12\,x\,Y + 4\,\left( 3 + x^2 \right) \,Y^2 + 
        8\,x\,Y^3 + 4\,Y^4 \right] \,{\rm erf}(x + Y) \right\} \ .
\label{FofxY}
\end{eqnarray}

\begin{figure}[ht]
\centering
\includegraphics[width=0.5\hsize]{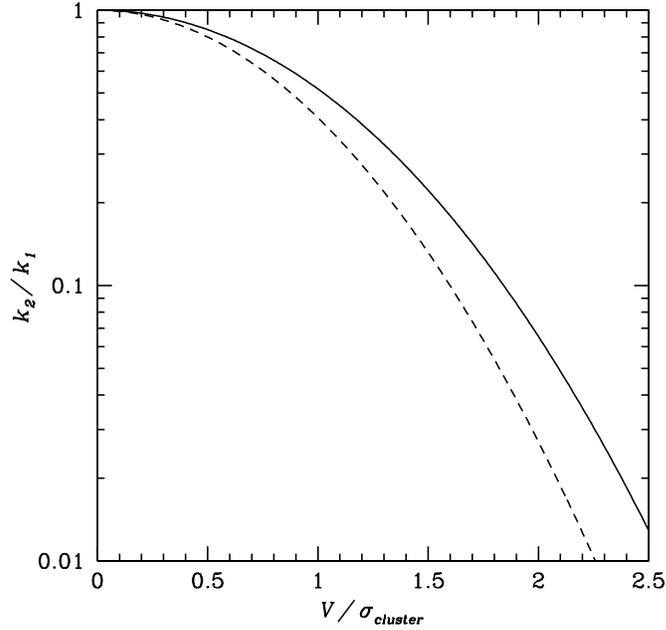}
\caption{Ratio of the rates of direct mergers of a galaxy in a cluster with
  velocity 
  dispersion $\sigma$ with galaxies of
another equal mass 
cluster passing right through the first cluster at velocity $V$ to
  that of mergers with galaxies of its own cluster (eqs.~[\ref{rat}] and
  [\ref{FofxY}]). 
The merger cross section (eq.~[\ref{Sofv}])  
involves a critical impact parameter that falls
  linearly with pericenter velocity, with a maximum velocity $v_0 =
  3.1\,\sqrt{3}\,\sigma_g$ (Roos \& Norman 1979), where $\sigma_g$ is the
  galaxy velocity dispersion, equal to 1/4 (\emph{solid curve}) or 1/8
  (\emph{dashed curve}) times the cluster velocity dispersion.
}
\label{ratfig}
\end{figure}
Figure~\ref{ratfig} displays $k_2/k_1$ as a function of $V/\sigma$ for two
values of $\sigma_g/\sigma$ for the linear critical parameter versus
velocity, with maximum merger velocity  $v_0 = 3.1\,\sqrt{3}\,\sigma_g$
\citep{RN79}. 
\twocolumn


\begin{thebibliography}{}

\bibitem[\protect\citeauthoryear{{Abell}, {Corwin}, \& {Olowin}}{{Abell}
  et~al.}{1989}]{ACO89}
{Abell} G.~O., {Corwin} H.~G., Jr.,  {Olowin} R.~P., 1989, \apjs, 70, 1

\bibitem[\protect\citeauthoryear{{Bahcall} \& {Chokshi}}{{Bahcall} \&
  {Chokshi}}{1992}]{Bahcall92}
{Bahcall} N.~A.,  {Chokshi} A., 1992, \apjl, 385, L33

\bibitem[\protect\citeauthoryear{{Bardelli}, {Zucca}, \& {Baldi}}{{Bardelli}
  et~al.}{2001}]{Bardelli01}
{Bardelli} S., {Zucca} E.,  {Baldi} A., 2001, \mnras, 320, 387

\bibitem[\protect\citeauthoryear{{Bardelli} et~al.}{{Bardelli}
  et~al.}{2000}]{Bardelli00}
{Bardelli} S., {Zucca} E., {Zamorani} G., {Moscardini} L.,  {Scaramella} R.,
  2000, \mnras, 312, 540

\bibitem[\protect\citeauthoryear{{Bardelli} et~al.}{{Bardelli}
  et~al.}{1998}]{Bardelli98}
{Bardelli} S., {Zucca} E., {Zamorani} G., {Vettolani} G.,  {Scaramella} R.,
  1998, \mnras, 296, 599

\bibitem[\protect\citeauthoryear{{Barnes}}{{Barnes}}{1988}]{Barnes88}
{Barnes} J.~E., 1988, \apj, 331, 699

\bibitem[\protect\citeauthoryear{{Becker}, {White}, \& {Helfand}}{{Becker}
  et~al.}{1995}]{Becker95}
{Becker} R.~H., {White} R.~L.,  {Helfand} D.~J., 1995, \apj, 450, 559

\bibitem[\protect\citeauthoryear{{Bekki} \& {Couch}}{{Bekki} \&
  {Couch}}{2003}]{Bekki03}
{Bekki} K.,  {Couch} W.~J., 2003, \apjl, 596, L13

\bibitem[\protect\citeauthoryear{{Best}}{{Best}}{2004}]{Best04}
{Best} P.~N., 2004, \mnras, 351, 70

\bibitem[\protect\citeauthoryear{{Best} et~al.}{{Best} et~al.}{2007}]{Best+07}
{Best} P.~N., {von der Linden} A., {Kauffmann} G., {Heckman} T.~M.,  {Kaiser}
  C.~R., 2007, \mnras, 379, 894

\bibitem[\protect\citeauthoryear{{Brand} et~al.}{{Brand}
  et~al.}{2003}]{Brand03}
{Brand} K., {Rawlings} S., {Hill} G.~J., et~al., 2003, \mnras, 344, 283

\bibitem[\protect\citeauthoryear{{Burns}}{{Burns}}{1990}]{Burns90}
{Burns} J.~O., 1990, \aj, 99, 14

\bibitem[\protect\citeauthoryear{{Carlberg}, {Yee}, \& {Ellingson}}{{Carlberg}
  et~al.}{1997}]{CYE97}
{Carlberg} R.~G., {Yee} H.~K.~C.,  {Ellingson} E., 1997, \apj, 478, 462

\bibitem[\protect\citeauthoryear{{Colless} et~al.}{{Colless}
  et~al.}{2001}]{Colless01}
{Colless} M., {Dalton} G., {Maddox} S., et~al., 2001, \mnras, 328, 1039

\bibitem[\protect\citeauthoryear{{Condon}}{{Condon}}{1992}]{Condon92}
{Condon} J.~J., 1992, \araa, 30, 575

\bibitem[\protect\citeauthoryear{{Condon} et~al.}{{Condon}
  et~al.}{1998}]{Condon98}
{Condon} J.~J., {Cotton} W.~D., {Greisen} E.~W., et~al., 1998, \aj, 115, 1693

\bibitem[\protect\citeauthoryear{{Coziol} et~al.}{{Coziol}
  et~al.}{1998}]{Coziol98}
{Coziol} R., {Ribeiro} A.~L.~B., {de Carvalho} R.~R.,  {Capelato} H.~V., 1998,
  \apj, 493, 563

\bibitem[\protect\citeauthoryear{{Cutri} et~al.}{{Cutri}
  et~al.}{2006}]{2MASSExpCat}
{Cutri} R., {Skrutskie} M., {Van Dyk} S., et~al., 2006, 2MASS Explanatory
  Supplement, Technical report, 2MASSExpSuppl

\bibitem[\protect\citeauthoryear{{Di Matteo} et~al.}{{Di Matteo}
  et~al.}{2007}]{dMCMS07}
{Di Matteo} P., {Combes} F., {Melchior} A.-L.,  {Semelin} B., 2007, \aap, 468,
  61

\bibitem[\protect\citeauthoryear{{Dressler}, {Thompson}, \&
  {Shectman}}{{Dressler} et~al.}{1985}]{Dressler85}
{Dressler} A., {Thompson} I.~B.,  {Shectman} S.~A., 1985, \apj, 288, 481

\bibitem[\protect\citeauthoryear{{Drinkwater} et~al.}{{Drinkwater}
  et~al.}{2004}]{Drink04}
{Drinkwater} M.~J., {Parker} Q.~A., {Proust} D., {Slezak} E.,  {Quintana} H.,
  2004, PASA, 21, 89

\bibitem[\protect\citeauthoryear{{Fabian}}{{Fabian}}{1991}]{Fabian91}
{Fabian} A.~C., 1991, \mnras, 253, 29P

\bibitem[\protect\citeauthoryear{{Fabian} et~al.}{{Fabian}
  et~al.}{1986}]{FANM86}
{Fabian} A.~C., {Arnaud} K.~A., {Nulsen} P.~E.~J.,  {Mushotzky} R.~F., 1986,
  \apj, 305, 9

\bibitem[\protect\citeauthoryear{{Gastaldello} et~al.}{{Gastaldello}
  et~al.}{2003}]{Gastaldello+03}
{Gastaldello} F., {Ettori} S., {Molendi} S., et~al., 2003, \aap, 411, 21

\bibitem[\protect\citeauthoryear{{Gavazzi} \& {Jaffe}}{{Gavazzi} \&
  {Jaffe}}{1986}]{GJ86}
{Gavazzi} G.,  {Jaffe} W., 1986, \apj, 310, 53

\bibitem[\protect\citeauthoryear{{Gerin}, {Combes}, \& {Athanassoula}}{{Gerin}
  et~al.}{1990}]{GCA90}
{Gerin} M., {Combes} F.,  {Athanassoula} E., 1990, \aap, 230, 37

\bibitem[\protect\citeauthoryear{{Gill}, {Knebe}, \& {Gibson}}{{Gill}
  et~al.}{2005}]{GKG05}
{Gill} S.~P.~D., {Knebe} A.,  {Gibson} B.~K., 2005, \mnras, 356, 1327

\bibitem[\protect\citeauthoryear{{Gunn} \& {Gott}}{{Gunn} \&
  {Gott}}{1972}]{GG72}
{Gunn} J.~E.,  {Gott} J.~R., 1972, \apj, 176, 1

\bibitem[\protect\citeauthoryear{{Hill} \& {Lilly}}{{Hill} \&
  {Lilly}}{1991}]{Hill91}
{Hill} G.~J.,  {Lilly} S.~J., 1991, \apj, 367, 1

\bibitem[\protect\citeauthoryear{{Jarrett} et~al.}{{Jarrett}
  et~al.}{2000}]{Jarrett00}
{Jarrett} T.~H., {Chester} T., {Cutri} R., et~al., 2000, \aj, 119, 2498

\bibitem[\protect\citeauthoryear{{Jones} et~al.}{{Jones}
  et~al.}{2004}]{Jones+04}
{Jones} D.~H., {Saunders} W., {Colless} M., et~al., 2004, \mnras, 355, 747

\bibitem[\protect\citeauthoryear{{Joseph} \& {Wright}}{{Joseph} \&
  {Wright}}{1985}]{JW85}
{Joseph} R.~D.,  {Wright} G.~S., 1985, \mnras, 214, 87

\bibitem[\protect\citeauthoryear{{Kaldare} et~al.}{{Kaldare}
  et~al.}{2003}]{Kaldare03}
{Kaldare} R., {Colless} M., {Raychaudhury} S.,  {Peterson} B.~A., 2003, \mnras,
  339, 652

\bibitem[\protect\citeauthoryear{{Ledlow} \& {Owen}}{{Ledlow} \&
  {Owen}}{1996}]{Ledlow96}
{Ledlow} M.~J.,  {Owen} F.~N., 1996, \aj, 112, 9

\bibitem[\protect\citeauthoryear{{{\L}okas} \& {Mamon}}{{{\L}okas} \&
  {Mamon}}{2001}]{LM01}
{{\L}okas} E.~L.,  {Mamon} G.~A., 2001, \mnras, 321, 155

\bibitem[\protect\citeauthoryear{{Machalski} \& {Godlowski}}{{Machalski} \&
  {Godlowski}}{2000}]{Machalski00}
{Machalski} J.,  {Godlowski} W., 2000, \aap, 360, 463

\bibitem[\protect\citeauthoryear{{Magliocchetti} et~al.}{{Magliocchetti}
  et~al.}{2004}]{Magliocchetti+04}
{Magliocchetti} M., {Maddox} S.~J., {Hawkins} E., et~al., 2004, \mnras, 350,
  1485

\bibitem[\protect\citeauthoryear{{Mamon}}{{Mamon}}{1992}]{Mamon92}
{Mamon} G.~A., 1992, \apjl, 401, L3

\bibitem[\protect\citeauthoryear{{Mamon}}{{Mamon}}{2000}]{Mamon00_IAP}
{Mamon} G.~A., 2000, in 15th IAP Astrophys.
  Mtg., Dynamics of Galaxies: from the early Universe to the Present, ed.
  F. {Combes}, G.~A. {Mamon} \& V. {Charmandaris}, Vol. 197.
\newblock (ASP, San Francisco), p. 377, arXiv:astro-ph/9911333

\bibitem[\protect\citeauthoryear{{Mamon} \& {{\L}okas}}{{Mamon} \&
  {{\L}okas}}{2005}]{ML05a}
{Mamon} G.~A.,  {{\L}okas} E.~L., 2005, \mnras, 362, 95

\bibitem[\protect\citeauthoryear{{Mamon} et~al.}{{Mamon} et~al.}{2004}]{MSSS04}
{Mamon} G.~A., {Sanchis} T., {Salvador-Sol{\' e}} E.,  {Solanes} J.~M., 2004,
  \aap, 414, 445

\bibitem[\protect\citeauthoryear{{Miller}}{{Miller}}{2005}]{Miller05}
{Miller} N.~A., 2005, \aj, 130, 2541

\bibitem[\protect\citeauthoryear{{Miller} \& {Owen}}{{Miller} \&
  {Owen}}{2001}]{MO01}
{Miller} N.~A.,  {Owen} F.~N., 2001, \apjs, 134, 355

\bibitem[\protect\citeauthoryear{{Miller} \& {Owen}}{{Miller} \&
  {Owen}}{2002}]{Miller02}
{Miller} N.~A.,  {Owen} F.~N., 2002, \aj, 124, 2453

\bibitem[\protect\citeauthoryear{{Miller} \& {Owen}}{{Miller} \&
  {Owen}}{2003}]{Miller03}
{Miller} N.~A.,  {Owen} F.~N., 2003, \aj, 125, 2427

\bibitem[\protect\citeauthoryear{{Monaco} et~al.}{{Monaco}
  et~al.}{1994}]{Monaco94}
{Monaco} P., {Giuricin} G., {Mardirossian} F.,  {Mezzetti} M., 1994, \apj, 436,
  576

\bibitem[\protect\citeauthoryear{{Navarro}, {Frenk}, \& {White}}{{Navarro}
  et~al.}{1996}]{NFW96}
{Navarro} J.~F., {Frenk} C.~S.,  {White} S.~D.~M., 1996, \apj, 462, 563

\bibitem[\protect\citeauthoryear{{Owen} et~al.}{{Owen} et~al.}{1999}]{Owen99}
{Owen} F.~N., {Ledlow} M.~J., {Keel} W.~C.,  {Morrison} G.~E., 1999, \aj, 118,
  633

\bibitem[\protect\citeauthoryear{{Peacock} \& {Nicholson}}{{Peacock} \&
  {Nicholson}}{1991}]{Peacock91}
{Peacock} J.~A.,  {Nicholson} D., 1991, \mnras, 253, 307

\bibitem[\protect\citeauthoryear{{Pietsch} et~al.}{{Pietsch}
  et~al.}{1997}]{PTAS97}
{Pietsch} W., {Trinchieri} G., {Arp} H.,  {Sulentic} J.~W., 1997, \aap, 322, 89

\bibitem[\protect\citeauthoryear{{Poggianti}}{{Poggianti}}{1997}]{Poggianti97}
{Poggianti} B.~M., 1997, \aaps, 122, 399

\bibitem[\protect\citeauthoryear{{Prestage} \& {Peacock}}{{Prestage} \&
  {Peacock}}{1988}]{Prestage88}
{Prestage} R.~M.,  {Peacock} J.~A., 1988, \mnras, 230, 131

\bibitem[\protect\citeauthoryear{{Proust} et~al.}{{Proust}
  et~al.}{2006}]{Proust+06}
{Proust} D., {Quintana} H., {Carrasco} E.~R., et~al., 2006, \aap, 447, 133

\bibitem[\protect\citeauthoryear{{Quintana}, {Carrasco}, \&
  {Reisenegger}}{{Quintana} et~al.}{2000}]{Quin00}
{Quintana} H., {Carrasco} E.~R.,  {Reisenegger} A., 2000, \aj, 120, 511

\bibitem[\protect\citeauthoryear{{Quintana} et~al.}{{Quintana}
  et~al.}{1995}]{Quin95}
{Quintana} H., {Ramirez} A., {Melnick} J., {Raychaudhury} S.,  {Slezak} E.,
  1995, \aj, 110, 463

\bibitem[\protect\citeauthoryear{{Raychaudhury}}{{Raychaudhury}}{1990}]{Raychaudhury90} 
{Raychaudhury} S., 1990, Ph.D. thesis, Univ. Cambridge

\bibitem[\protect\citeauthoryear{{Raychaudhury} et~al.}{{Raychaudhury}
  et~al.}{1991}]{Ray91}
{Raychaudhury} S., {Fabian} A.~C., {Edge} A.~C., {Jones} C.,  {Forman} W.,
  1991, \mnras, 248, 101

\bibitem[\protect\citeauthoryear{{Reisenegger} et~al.}{{Reisenegger}
  et~al.}{2000}]{Reisen00}
{Reisenegger} A., {Quintana} H., {Carrasco} E.~R.,  {Maze} J., 2000, \aj, 120,
  523

\bibitem[\protect\citeauthoryear{{Roettiger}, {Burns}, \& {Loken}}{{Roettiger}
  et~al.}{1993}]{RBL93}
{Roettiger} K., {Burns} J.,  {Loken} C., 1993, \apjl, 407, L53

\bibitem[\protect\citeauthoryear{{Roos}}{{Roos}}{1981}]{Roos81}
{Roos} N., 1981, \aap, 104, 218

\bibitem[\protect\citeauthoryear{{Roos} \& {Norman}}{{Roos} \&
  {Norman}}{1979}]{RN79}
{Roos} N.,  {Norman} C.~A., 1979, \aap, 76, 75

\bibitem[\protect\citeauthoryear{{Sadler} et~al.}{{Sadler}
  et~al.}{2002}]{Sadler02}
{Sadler} E.~M., {Jackson} C.~A., {Cannon} R.~D., et~al., 2002, \mnras, 329, 227

\bibitem[\protect\citeauthoryear{{Sanderson}, {Ponman}, \&
  {O'Sullivan}}{{Sanderson} et~al.}{2006}]{SPO06}
{Sanderson} A.~J.~R., {Ponman} T.~J.,  {O'Sullivan} E., 2006, \mnras, 372, 1496

\bibitem[\protect\citeauthoryear{{Strauss} et~al.}{{Strauss}
  et~al.}{2002}]{Strauss02}
{Strauss} M.~A., {Weinberg} D.~H., {Lupton} R.~H., et~al., 2002, \aj, 124, 1810

\bibitem[\protect\citeauthoryear{{Tully}}{{Tully}}{1988}]{Tully88}
{Tully} R.~B., 1988, \aj, 96, 73

\bibitem[\protect\citeauthoryear{{van der Hulst} \& {Rots}}{{van der Hulst} \&
  {Rots}}{1981}]{vdHR81}
{van der Hulst} J.~M.,  {Rots} A.~H., 1981, \aj, 86, 1775

\bibitem[\protect\citeauthoryear{{Venturi} et~al.}{{Venturi}
  et~al.}{1997}]{Venturi97}
{Venturi} T., {Bardelli} S., {Morganti} R.,  {Hunstead} R.~W., 1997, \mnras,
  285, 898

\bibitem[\protect\citeauthoryear{{Venturi} et~al.}{{Venturi}
  et~al.}{2000}]{Venturi00}
{Venturi} T., {Bardelli} S., {Morganti} R.,  {Hunstead} R.~W., 2000, \mnras,
  314, 594

\bibitem[\protect\citeauthoryear{{Venturi} et~al.}{{Venturi}
  et~al.}{2001}]{Venturi01}
{Venturi} T., {Bardelli} S., {Zambelli} G., {Morganti} R.,  {Hunstead} R.~W.,
  2001, \mnras, 324, 1131

\bibitem[\protect\citeauthoryear{{Yun}, {Reddy}, \& {Condon}}{{Yun}
  et~al.}{2001}]{Yun01}
{Yun} M.~S., {Reddy} N.~A.,  {Condon} J.~J., 2001, \apj, 554, 803

\end{thebibliography}
\end{document}